\documentclass{article}
\usepackage{fullpage,times,color,amsmath,amssymb,epsfig,subfigure,comment,graphicx,algorithm,algorithmic,cite,multirow,enumerate}
\usepackage{url}
\usepackage{multicol}
\usepackage{epstopdf}

\newtheorem{example}{Example}
\newtheorem{mydef}{Definition}
\newtheorem{observation}{Observation}

\newcommand{\scb}{SC$_{\textrm{b}}$}
\newcommand{\scp}{SC$_{\textrm{p}}$}
\newcommand{\tscp}{TSC$_{\textrm{p}}$}
\newcommand{\scg}{SC$_{\textrm{g}}$}

\makeatletter
\newcommand{\setalglineno}[1]{%
  \setcounter{ALC@line}{\numexpr#1-1}}
\makeatother

\begin{document}


\begin{center}
\vspace*{5cm}
{\huge\bf\sf Lightweight and Secure Two-Party Range Queries\\over Outsourced Encrypted Databases}\\
\vspace{3cm}

{\Large\bf\sf Bharath K. Samanthula$^1$, Wei Jiang$^{2}$ and Elisa Bertino$^{3}$}\\


\vspace{3cm}
{\Large\sf \today}\\ 
\vspace{1cm}
\hrule
\begin{center}
{\large\sf Technical Report \\
$^{1,3}$Department of Computer Science, Purdue University\\
305 N. University Street, West Lafayette, IN  47907\\
\{bsamanth, bertino\}@purdue.edu\\
$^{2}$Department of Computer Science, Missouri S\&T\\
500 W. 15th Street, Rolla, Missouri 65409\\
wjiang@mst.edu}\\
\end{center}

\end{center}

\title{}
\author{}


\date{}

\maketitle

\begin{abstract}
With the many benefits of cloud computing, an entity may want to 
outsource its data and their related analytics tasks to a cloud. 
When data are sensitive, it is in the interest of the entity to outsource
encrypted data to the cloud; however, this
limits the types of operations 
that can be performed on the cloud side. Especially, evaluating  
queries over the encrypted data stored on the cloud without the entity performing
any computation and without ever decrypting the data become a very challenging problem. In this paper, 
we propose solutions to conduct range queries  
over outsourced encrypted data. The existing methods leak valuable information to the cloud
which can violate the security guarantee of the underlying encryption schemes. 
In general, the main 
security primitive used to evaluate 
range queries is secure comparison (SC) of encrypted integers. However, we observe that 
the existing SC protocols are not very efficient. 
To this end, we first propose a novel SC scheme that takes encrypted integers and outputs 
encrypted comparison result.  We empirically show its 
practical advantage over the current state-of-the-art. We then utilize the proposed SC scheme to 
construct two new secure range query protocols. 
Our protocols protect data confidentiality, privacy of user's query, and also preserve 
the semantic security of the encrypted data; 
therefore, they are more secure than the existing protocols. Furthermore, 
our second protocol is lightweight at the user end, and it can 
allow an authorized user to use any device with limited storage and computing capability
 to perform the range queries over outsourced encrypted data.

\end{abstract}


\section{Introduction}\label{sec:intro}
For many companies, especially in 
the case of small and 
medium size businesses, maintaining their 
own data can be a challenging issue due to large capital expenditures and  
high day-to-day operational costs. Therefore, data owners may be more interested in 
outsourcing their data and operations related to the data. 
Along this direction, cloud computing \cite{qian-2009,chow-2009,armbrust-2010}
offers a promising solution due to various advantages such as 
cost-efficiency and flexibility. 
%
Due to various 
privacy reasons\cite{pearson-2009,pearson-2010,samarati-2010,takabi-2010} and as the cloud may not be fully 
trusted, users encrypt their data at first place and then outsource them to 
the cloud. However, this places limitations 
on the range of operations that can be performed 
over encrypted data in the cloud. In recent years, query processing over encrypted data stored in the cloud 
has gained significant importance as it is a common feature in many 
outsourced service-oriented databases. 

In this paper, we focus on 
processing range queries over encrypted data in the cloud. A 
range query, where records are retrieved if the values of a 
specific field lie in the range $(\alpha, \beta)$, 
is one among the highly desirable queries. For example, consider the situation 
where a hospital outsources its patients' medical data to a cloud after the data 
were properly encrypted. If at some future 
time, suppose a researcher wants to access this hospital's data for analyzing 
the disease patterns of all the 
young patients whose ages lie between 18 and 25. For privacy reasons, the input query by the 
researcher should not be 
revealed to the cloud. In addition, due to efficiency and privacy reasons, the entire patients' medical data 
should not be revealed to the researcher. That is, on one hand, we claim that a trivial solution where an 
authorized user can download the whole data from the cloud and decrypt them 
to perform range query locally is not practical from user's computation perspective. 
On the other hand, for privacy reasons, only the disease information of a patient whose age is in (18, 25) should 
be revealed to the researcher. We refer to such a process as 
privacy-preserving range query (PPRQ) over encrypted data. At a high level, the  
PPRQ protocol should securely compare the user's search input (i.e., $\alpha$ and $\beta$) with 
the encrypted field values (stored in the cloud) 
upon which the user wants to filter the data records. 

Based on the above discussions, it is clear that 
the underlying basic security primitive required to solve 
the PPRQ problem is secure comparison of encrypted integers. 
Secure comparison (SC) is an important building block in many 
distributed and privacy-preserving applications 
such as secure electronic voting (e.g., \cite{clarkson-2008}), private auctioning and 
bidding (e.g., \cite{cachin-1999,bogetoft-2009}), and privacy-preserving data 
mining (e.g., \cite{agrawal2000privacy,lindell2000privacy}). First, we observe that 
the existing custom-designed SC protocols (e.g., \cite{garay-2007,damgaard-2007,blake-2009}) require encryptions of individual bits 
of inputs rather than simple encrypted integers; therefore, making them less efficient. Secondly, the traditional two-party computation 
methods based on Yao's garbled-circuit technique seem to a better choice to solve the SC problem. Indeed, some recent 
implementations, such as FastGC\cite{huang-2011}, demonstrate that such generic approaches can 
outperform the custom-designed protocols. Nevertheless, we show that the SC protocol constructed 
using Yao's garbled-circuit on FastGC is still less efficient (see Section \ref{sec:new-sc} for details). 
Along this direction, we first 
propose a novel SC protocol that is more efficient than the methods based on the 
above two approaches.  

Apart from ensuring data confidentiality, which is commonly achieved by encrypting 
the data before outsourcing, two important privacy issues 
related to the PPRQ problem are: 1) preserving 
the privacy or confidentiality of an input query and 2) preventing 
the cloud from learning the data access patterns. While 
the privacy of user's input query can be protected by the security of the underlying 
encryption schemes, hiding the 
data access patterns from the cloud is a challenging task. This is because of the fact that the 
encrypted data resides in the cloud (which acts as a 
third party). As mentioned in \cite{vimercati-2012,williams-2008}, by monitoring 
the data access patterns, the cloud can reconstruct the correspondence between 
the plaintext data and the encrypted data based on the access pattern frequencies to each piece 
of data. These access patterns can actually violate the security guarantee of 
the underlying encryption schemes used to encrypt the outsourced data. 

\subsection{Access Patterns and Semantic Security}\label{subsec:access-patterns}
By data access patterns, we mean the relationships among the encrypted data 
that can be observed by the cloud during query processing. For example,
suppose there are five records $t_1, \ldots, t_5$ in a database $D$ and let 
$E_{pk}(.)$ denote a semantically secure encryption function, such as Paillier 
cryptosystem\cite{paillier-99}. Assume that these records
are encrypted $($i.e., $E_{pk}(t_1), \ldots, E_{pk}(t_5))$ and stored on a cloud,
denoted by $C$. After processing a user's range query, let $E_{pk}(t_2)$ and $E_{pk}(t_5)$ be the
output returned to the user. More details on how this is achieved are given in 
Section \ref{sec:proposed}. 

In cryptography, it is a common belief that an encryption scheme 
needs to be at least secure against chosen-plaintext attack (i.e., semantic security). 
In other words, the ciphertexts should be indistinguishable from an (computationally 
bounded) adversary's perspective.
From the previous example, before processing the user query, $C$ cannot distinguish
$E_{pk}(t_1), \ldots, E_{pk}(t_5)$ because $E_{pk}(.)$ is semantically secure. However, after 
processing the user query, $C$ learns that the encrypted data
 can be partitioned into two groups: $\{E_{pk}(1), E_{pk}(3), E_{pk}(4)\}$ and  
$\{E_{pk}(2), E_{pk}(5)\}$. More specifically, $\{E_{pk}(1), E_{pk}(3), E_{pk}(4)\}$ is distinguishable from 
$\{E_{pk}(2),$ $E_{pk}(5)\}$. This breaks the semantic security of $E_{pk}(.)$. Thus, to 
protect the confidentiality or semantic security 
of the outsourced data, access patterns should be hidden
from the cloud who stores and processes  the data. 
 
A naive approach to hide the access patterns is to encrypt the database with symmetric key encryption 
schemes (e.g., AES) and then outsourcing 
them to the cloud. However, 
during the query processing step, the cloud cannot perform any algebraic operations 
over the encrypted data. 
Thus, the entire encrypted database has to be downloaded by the authorized user which 
is not practical especially for mobile users and large databases. 
On the other hand, to avoid downloading the entire encrypted database from the cloud and 
to hide access patterns, 
a user can adopt Oblivious RAM (ORAM) techniques \cite{williams-2008,shi-2011}. 

The main goal of Oblivious
RAM is to hide which data record has been accessed by the user. 
To utilize ORAM, the user needs to know where to retrieve the record on the
cloud through certain indexing structure. Since a dataset can contain hundreds of attributes and 
a range query can be performed on any of these attributes, one indexing structure to utilize ORAM
is clearly insufficient. How to efficiently utilize multiple indexes on multiple attribute of the data is 
still an open problem with ORAM techniques. In addition, since the current computations required for ORAM 
techniques cannot be  parallelized, we cannot take full advantage  of the  
 large-scale parallel processing capability of a cloud.   

To provide better security guarantee and shift the entire computation to the
cloud, 
this paper 
proposes two novel PPRQ protocols by utilizing our new SC scheme as the building block. Our protocols 
protect data confidentiality and privacy of user's input query. At the same time, they hide data access patterns 
from the cloud service providers. Also, our second 
protocol is very efficient from the end-user perspective.
\subsection{Our Contributions}
We propose efficient protocols for secure comparison and 
PPRQ problems over encrypted data. More specifically, the main 
contributions of this paper are two-fold: 
\begin{enumerate}[(i).]
\item \textit{Secure Comparison.} As mentioned earlier, the basic security primitive required 
to solve the PPRQ problem is secure comparison (SC) of encrypted integers. 
Since the existing SC methods are not that efficient, we first propose 
an efficient and probabilistic SC scheme. 
Because the proposed SC scheme is probabilistic in nature, we theoretically analyze 
its correctness and provide a formal security proof based 
on the standard simulation paradigm\cite{goldreich-general}. We stress that, our SC scheme returns 
the correct output for all practical applications. 
%
\item  
\textit{Privacy-Preserving Range Query.} We construct two novel 
PPRQ protocols using our new SC scheme as the building block. Our 
protocols achieve the desired security objectives of PPRQ (see Section \ref{sec:probstat-threatmodel} for 
details), and the computation cost on the end-user is very low since most computations are shifted to the cloud.
Also, the 
computations performed by the cloud can be easily parallelized to drastically 
improve the response time.
\end{enumerate}
The rest of the paper is organized as follows. 
In Section \ref{sec:probstat-threatmodel}, we formally discuss our problem statement along 
with the threat model adopted in this paper.  
A brief survey 
of the existing related work is presented in Section \ref{sec:prev-work}.
Our new secure comparison scheme is presented in Section \ref{sec:new-sc} along 
with a running example. Also, apart from providing a formal security proof, we theoretically 
analyze the accuracy guarantee of our SC scheme. Additionally, in this section,  
we empirically compare the performance of our SC scheme with the existing methods and demonstrate 
its practical applicability. 
The proposed PPRQ protocols, which are constructed on the top of our SC scheme, are 
presented in Section \ref{sec:proposed}. 
Finally, we conclude the paper with possible future work in Section \ref{sec:concl}.

\section{Problem Settings and Threat Model}\label{sec:probstat-threatmodel}
\subsection{Architecture and Desired Security Properties}
Consider a data owner Alice holding the database $D$ with 
$n$ records denoted by $\langle t_1, \ldots, t_n\rangle$. Let $t_{i,j}$ denote the 
$j^{th}$ attribute value of tuple $t_i$, for $1 \leq i \leq n$ and $1 \leq j \leq w$, 
where $w$ denotes the number of attributes. We assume that Alice encrypts her 
database $D$ attribute-wise with an additive homomorphic 
encryption scheme that is semantically secure, such as Paillier cryptosystem \cite{paillier-99},  
and outsources the encrypted database to the cloud. Without loss of generality, let $T$ denote the encrypted 
database. Besides the data, Alice outsources 
the future query processing services to the cloud. Now, consider a user Bob who is authorized by Alice  
to access $T$ in the cloud. Suppose, if at some future time, 
Bob wants to execute a range query $Q = \{k, \alpha, \beta\}$ over encrypted data 
in the cloud, where $k$ is the attribute index upon which 
he wants to filter the records with $\alpha$ and $\beta$ as the lower 
and upper bound values, respectively. Briefly, the goal 
of the PPRQ protocol is to securely retrieve the set of records, denoted by $S$, such that 
the following property holds.
$$\forall~t'\in S, \alpha \le t'_{k} \le \beta$$
where $t'_k$ denotes the $k^{th}$ attribute value of data record $t'$. More formally, we define the PPRQ protocol as follows:
$$\textrm{PPRQ}(T, Q) \rightarrow S$$
For any given PPRQ protocol, we stress that the following privacy requirements should be met:
\begin{enumerate}[(a)]
\item Bob's input query $Q$ should not be revealed to the cloud. 
\item During any of the query processing steps, contents of $D$ should not be disclosed to the cloud.
\item The data access patterns should not be revealed to the cloud. That is, for any given input 
query $Q$, the cloud should not know which data records in $D$ belong 
to the corresponding output set $S$. Also, access patterns related to 
any intermediate computations should not be revealed to the cloud. In other
words, the semantic security of the encrypted data needs to be preserved. 
\item $D-S$ (i.e., the set of records not satisfying $Q$) should not be disclosed to Bob.
\item At the end of the PPRQ protocol, $S$ should be revealed only to Bob and no 
information is revealed to the cloud. 
\end{enumerate}
\subsection{Threat Model}\label{subsec:secure-def}

In this paper, 
privacy/security is closely related to the 
amount of information disclosed during the execution of a protocol. 
Proving the security of a distributed protocol is very different from 
that of an encryption scheme. In the proposed protocols, our goal is to ensure no information leakage 
to the participating parties other than what they can deduce from 
their own inputs and outputs. To maximize security guarantee, 
we adopt the commonly accepted security definitions and proof techniques in the literature 
of secure multiparty computation (SMC) to analyze the security of the proposed protocols. 
SMC was first introduced by Yao's Millionaires' (two-party) problem \cite{yao-82,yao-86}, and
it was extended by Goldreich et al. \cite{Goldreich87} to the multi-party case. It was proved 
in \cite{Goldreich87} that any computation which can be done in polynomial time by a single party can also be 
done securely by multiple parties. 


There are three common adversarial models under SMC: semi-honest, covert and malicious.
An adversarial model generally specifies what an adversary or attacker is allowed to do
during an execution of a secure protocol. In 
the semi-honest model, an attacker (i.e., one of the participating parties) 
is expected to follow the prescribed steps of a protocol. However,
the attacker can compute any additional information based on his or her  private input, output and messages
received during an execution of the secure protocol. 
As a result, whatever can be inferred from the 
private input and output of an attacker is not considered as a privacy violation. 
An adversary in the semi-honest model can be treated as a passive attacker whereas 
an adversary in the malicious model can be treated as an active attacker who can arbitrarily diverge from 
the normal execution of a protocol. On the other hand, the covert adversary model\cite{aumann-2010}  lies 
between the semi-honest and malicious model. More specifically, 
an adversary under the covert model may deviate arbitrarily 
from the rules of a protocol, however, in the case of cheating, 
the honest party is guaranteed to detect this cheating with good probability.  

In this paper, to develop secure and efficient protocols,
we assume that parties are semi-honest for two reasons. First, as mentioned in \cite{huang-2011}, 
developing protocols under the semi-honest setting is an important first step 
towards constructing protocols with stronger security guarantees. Second, it is worth pointing out that most 
practical SMC protocols proposed in the 
literature (e.g., \cite{henecka-2010,PSI-NDSS,huang-2011,nikolaenko-2013}) are implemented 
only under the semi-honest model. By semi-honest model, we implicitly  
assume that the cloud service providers (or other participating users) 
utilized in our protocols do not collude. Since current known cloud service
providers are well established IT companies, it is hard to see the possibility
for two companies, e.g., Google and Amazon, to collude to damage their
reputations and consequently place negative impact on their revenues.
Thus, in our problem domain, assuming the participating parties are
semi-honest is very realistic. 

However, in Section \ref{security-scp}(2), we discuss strategies to extend the proposed SC protocol to be secure
under the malicious and the covert models. Since SC is the main component of the proposal PPRQ protocols,
we believe that the same strategies can be used to make the PPRQ protocols secure under the malicious 
and covert models. Due to space limitations, we will not provide detailed discussions on how to modify the 
proposed PPRQ protocols to be secure under other adversarial models and leave it as part of our future work. 

Formally, the following definition captures the security of a protocol under the semi-honest model \cite{goldreich-general}.
\begin{mydef}\label{def:semi-honest}
Let $a_i$ be the input of party $P_i$, $\Pi_i(\pi)$ be  $P_i$'s 
execution image of the protocol $\pi$ and $b_i$ be its output computed from $\pi$.
Then, $\pi$ is secure if $\Pi_i(\pi)$ can be simulated from
$\langle a_i, b_i \rangle$ and
distribution of the simulated image is computationally indistinguishable
from $\Pi_i(\pi)$.
\end{mydef} 

In the above definition, an execution image generally includes the input, the output and
the messages communicated during an execution of a protocol. To prove 
a protocol is secure under the semi-honest model, we generally need to show that the execution image of a protocol 
does not leak any information regarding the private inputs of 
participating parties \cite{goldreich-general}. 

\section{Related Work}\label{sec:prev-work}
In this section, we first briefly review upon the existing work 
related to our problem domain. Then, we refer to the  
additive homomorphic properties and the corresponding encryption scheme adopted in 
this paper. Finally, we discuss the secure comparison problem and point out 
two (different) best-known solutions to solve this problem. 

\subsection{Keyword Search on Encrypted Data}
A different but closely related work to querying on encrypted data is 
``keyword search on encrypted data''. The main goal of this problem is to retrieve 
the set of encrypted files stored on a remote server (such as the cloud) that match 
the  user's input keywords. Along this direction, much work has been published 
based on searchable encryption schemes 
(e.g., \cite{wang-2010,yang-2011,cao-2011,blass-2012}). However, these works mostly concentrate 
on protecting data confidentiality and they do not protect data 
access patterns. Though some recent works addressed the issue 
of protecting access patterns while searching for 
keywords\cite{islam-2012,kuzu-2012}, at this point, it is not clear 
how their work can be mapped to range queries which is an entirely different and complex problem than 
simple exact matching.

\subsection{Existing PPRQ Methods}
The PPRQ problem has been investigated under 
different security models such as order-preserving encryption \cite{agrawal2004order,boldyreva-2009} and searchable 
public key encryption schemes\cite{boneh-2007,shi-2007}. Range queries over 
encrypted data was first 
addressed by Agrawal et al.\cite{agrawal2004order}. They have developed an order-preserving 
encryption (OPE) scheme for numeric data that can support indexing to efficiently 
access the encrypted data stored on an untrusted server. The basic idea behind the 
OPE scheme is to map plaintexts into ciphertexts by preserving their relative order. That is, 
for any given ciphertexts $c_1$ and $c_2$ corresponding to plaintexts 
$p_1$ and $p_2$, if $p_1 \ge p_2$ then it is guaranteed that $c_1 \ge c_2$. Such 
a guarantee allows the untrusted server (i.e., the cloud in our case) to easily process range queries 
even if the data are encrypted. 
As an improvement, Boldyreva et al. \cite{boldyreva-2009} provided a formal security 
analysis and an efficient version of the OPE scheme. Nevertheless, the main 
disadvantage of OPE schemes is that they are not secure against chosen-plaintext attacks (CPA). 
This is because of the fact that OPE schemes are deterministic (i.e., different encryptions 
of a given plaintext will result in the same ciphertext) and they reveal relative ordering  
among plaintexts. Due to the above reasons, the ciphertexts are distinguishable from the server's perspective; therefore, 
OPE schemes are not IND-CPA secure. 

As an alternative, in the past few years, researchers have been focusing on 
searchable public key encryption schemes by leveraging cryptographic techniques. Along 
this direction, in particular to range queries, some earlier works 
\cite{boneh-2007,shi-2007} were partly successful in addressing the PPRQ problem. However, as 
mentioned in \cite{hore-2012}, these methods are susceptible to value-localization problem; 
therefore, they are not secure. In addition, they leak data access patterns to the server. 
Recently, Hore et al.\cite{hore-2012} developed 
a new multi-dimensional PPRQ protocol by securely generating 
index tags for the data using bucketization techniques. However, their method is susceptible to access 
pattern attacks (this issue was also mentioned as a 
drawback in \cite{hore-2012}) and false positives in the 
returned set of records. More specifically, the final set of records has to be weeded by the client to remove 
false positives which incurs computational overhead on 
the client side. In addition, since the bucket labels are revealed 
to the server, we believe that their method may lead to 
unwanted information leakage. 

Vimercati et al.\cite{vimercati-2011} proposed a new technique for protecting confidentiality 
as well as access patterns to the data in outsourced environments. Their technique 
is based on constructing shuffled index structures using B+-trees. In order to hide 
the access patterns, their method introduces fake searches in conjunction with 
the actual index value to be searched. We emphasize that their work solves a different 
problem - mainly how to securely outsource the index and then obliviously search over 
this data structure. Their technique has a straight-forward application to keyword search 
over encrypted data since it deals with exact matching. However, at this point, it is not clear how their work 
can be extended for range queries that require implicit comparison operations to be 
performed in a secure manner.

We may ask if 
we can use fully homomorphic cryptosystems (e.g., \cite{gentry-2009}) which can 
perform arbitrary computations over encrypted data without ever decrypting them. 
However, such techniques are very expensive and their usage in practical 
applications have yet to be explored. For example, it was shown in \cite{gentry-2011} that even 
for weak security parameters one ``bootstrapping'' operation of the homomorphic operation 
would take at least 30 seconds on a high performance machine. 

As an independent work, Bajaj et al.\cite{bajaj-2011} developed a new prototype 
to execute SQL queries by leveraging server-hosted tamper-proof trusted hardware in 
critical query processing stages. However, their work still 
reveals data access patterns to the server. Recently, Samanthula et al.\cite{bksam-cloud13} 
proposed a new PPRQ protocol by utilizing the secure comparison (SC) protocol in \cite{blake-2009} 
as the building block. Perhaps, their method is the most 
closely related work to the protocols proposed in this paper. 
However, the SC protocol in \cite{blake-2009} operates 
on encrypted bits rather than on encrypted integers; therefore, the overall throughput in their protocol 
is less. In addition, their protocol leaks data  access patterns to the cloud service provider. 

Hence, in order to provide better 
security and improve efficiency, this paper proposes two novel PPRQ protocols that protect 
data confidentiality, privacy of user's input query and hide data access patterns. 

\subsection{Additive Homomorphic Encryption Scheme}
In the proposed protocols, we utilize an additive homomorphic 
encryption scheme (denoted by HEnc$^+$) that is probabilistic in nature. Without 
loss of generality, let $E_{pk}$ and $D_{sk}$ be the encryption and decryption 
functions of an HEnc$^+$ system, where $pk$ and $sk$ are the public and secret 
keys, respectively. Given a ciphertext and $pk$, it is 
impossible for an (computationally bounded) adversary 
to retrieve the corresponding plaintext in polynomial time. Let $N$ denote 
the RSA modulus (or part of public key $pk$). In general, the HEnc$^+$ 
system exhibits the following properties:
\begin{itemize}
\item Given two ciphertexts $E_{pk}(a)$ and $E_{pk}(b)$, where $a,b \in \mathbb{Z}_N$, we can compute 
the ciphertext corresponding to $a+b$ by performing homomorphic 
addition (denoted by $+_h$) on the two ciphertexts: $$D_{sk}(E_{pk}(a)+_h E_{pk}(b)) = a+b;$$
\item Using the above property, for any given constant $u \in \mathbb{Z}_N$, the homomorphic multiplication property is given by:
$$D_{sk}(E_{pk}(a)^u) = a\ast u;$$
\item The encryption scheme is semantically secure\cite{goldreich-encryption}, 
that is indistinguishability under chosen-plaintext attack (IND-CPA) holds.  
\end{itemize}
Any HEnc$^+$ system can be used to implement the proposed protocols; however, 
this paper uses the Paillier cryptosystem\cite{paillier-99} due to its efficiency. 

\subsection{Secure Comparison (SC)}\label{subsec:sc-disc}
Let us consider a party $P_1$ holding two Paillier encrypted values $(E_{pk}(x), E_{pk}(y))$ and a party 
$P_2$ holding the secret key $sk$ such that $(x, y)$ is unknown to both 
parties. The goal of the secure comparison (SC) protocol is for $P_1$ and $P_2$ 
to securely evaluate the functionality $x \ge y$. The comparison result, denoted by $c$, is 1 if $x\ge y$, and 0 otherwise.  
At the end of the SC protocol, the output $E_{pk}(c)$ should be known only to $P_1$. During 
this process, no other information regarding $x, y,$ and $c$ is revealed to $P_1$ and $P_2$. 

We emphasize that other variations 
of SC include $(x, E_{pk}(y))$, $(E_{pk}(x), y)$, or shares of $x$ 
and $y$ as private inputs. On one hand, the existing SC methods based on 
Yao's garbled-circuit technique (e.g., \cite{kolesnikov-2009}) 
assume that $x$ and $y$ are known to $P_1$ and $P_2$ respectively. However, 
such techniques can be easily modified to handle the above input cases  
with minimal cost. For completeness, here we briefly explain how to construct a secure 
comparison circuit (denoted by \scg) using 
$E_{pk}(a)$ and $E_{pk}(b)$ as $P_1$'s input. Since it would be complex (and costly) to include encryption and 
decryption operations as a part of the circuit, we discuss a simple method to compute the random shares of $x$ 
and $y$ from $(E_{pk}(x), E_{pk}(y))$ using homomorphic properties. Initially, $P_1$ masks
the encrypted inputs by computing $E_{pk}(x+ r_1)$ and $E_{pk}(y+r_2)$, where  $r_1$ and $r_2$ are 
random numbers (known only to $P_1$) in $\mathbb{Z}_N$, and sends them to $P_2$. Upon decryption, $P_2$ gets 
his/her random shares as $x + r_1 \bmod N$ and $y+r_2\bmod N$. Also, $P_1$ sets his/her 
corresponding random shares as $N - r_1$ and $N- r_2$. After this, $P_1$ can 
construct a garbled-circuit, where $P_2$ acts as the circuit evaluator, based on the following steps:
\begin{enumerate}[(i).]
\item Add the random shares (as a part of the circuit) to get $x$ and $y$. 
\item Compute the comparison result on $x$ and $y$ using \cite{kolesnikov-2009}. We 
stress that the comparison result $c$ is not known to either of the parties since the 
result is encoded as a part of the garbled-circuit.
\item Add a random value (known only to $P_1$) to the comparison result. The masked 
value is the final output of the circuit which will be known only to $P_2$. 
\end{enumerate}
Next, $P_2$ encrypts the masked comparison result and sends it to $P_1$. Finally $P_1$ removes 
the masking factor to compute 
$E_{pk}(c)$ using homomorphic properties. We emphasize that the addition operations 
in the above circuit should be followed by an implicit modulo $N$ operation to compute 
the correct result. At a high level, the above circuit seems to be simple and efficient. 
Nevertheless, as we show in Section \ref{subsec:sc-exp}, such traditional techniques are much  
less efficient than our proposed SC scheme. 

In this paper, we do not consider the existing secure comparison protocols that 
are secure under the information theoretic setting. This is because, the existing secure 
comparison protocols under the information theoretic setting are commonly based on 
linear secret sharing schemes, such as Shamir's\cite{shamir-1979}, which require at least 
three parties. We emphasize that our problem setting is entirely different than those methods since the data 
in our case are encrypted and our protocols require only two parties. Our protocols, which are 
based on additive homomorphic encryption schemes, are orthogonal to the secret sharing based SC schemes. 
Nevertheless, developing a PPRQ protocol by using the secret sharing based SC methods that protect the 
data access patterns is still an open problem; therefore, it can be treated as 
an interesting future work.

On the other hand, there exist a large number of custom-designed 
SC protocols (e.g., \cite{garay-2007,damgaard-2007,blake-2009}) that directly 
operate on encrypted inputs. 
Since the goal of this paper is not to investigate all the existing SC protocols, 
we simply refer to the most efficient known implementation of SC (here we consider methods 
based on Paillier cryptosystem to have 
a fair comparison with our scheme) that was proposed 
by Blake et al.\cite{blake-2009}. We emphasize that the SC protocol given in \cite{blake-2009} requires 
the encryptions of individual bits of $x$ and $y$ as the input rather than $(E_{pk}(x), E_{pk}(y))$. Though 
their protocol is efficient than the above garbled-circuit based SC method (i.e., \scg) for smaller input domain sizes, 
we show that our SC scheme outperforms both the methods for all practical values of input domain 
sizes (see Section \ref{subsec:sc-exp} for details). Also, it is worth pointing out  
that the protocol in \cite{blake-2009} 
leaks the comparison result $c$ to at least one of the involved parties. However, 
by using the techniques in \cite{damgaard-2007}, we can easily modify (at the expense of 
extra cost) the protocol of \cite{blake-2009} 
to generate $E_{pk}(c)$ as the output without revealing $c$ to both parties. 

\section{The Proposed SC Scheme}\label{sec:new-sc}
As mentioned above, which we also show empirically in 
the later part of this section, the existing well-known SC methods in \cite{blake-2009,kolesnikov-2009} 
are not that efficient. 
Therefore, to improve efficiency without compromising security, we propose a novel secure scheme, denoted by \scp, 
for efficient comparison of encrypted integers. In our \scp~protocol, the output is $E_{pk}(c)$ and is revealed 
only to $P_1$. That is, the comparison result $c$ is not revealed to $P_1$ and $P_2$. We stress that 
\scp~ can be easily modified to generate shared output. Therefore, depending on 
the application requirements, our \scp~protocol can be used as a building block in larger  
privacy-preserving tasks.

\begin{algorithm}[!thbp]
\begin{algorithmic}[1]
\REQUIRE $P_1$ has Paillier encrypted values $(E_{pk}(x), E_{pk}(y))$, where $(x , y)$ 
is not known to both parties and $0 \leq x,y < 2^m$;
	(Note: The public key $(g,N)$ is known to both parties whereas the 
secret key $sk$ is known only to $P_2$)\\
\STATE $P_1$: 
    \begin{enumerate}\itemsep=0pt 
          \item[(a).] $l \gets 2^{-1} \mod N$ and $d' \gets 0$
          \item[(b).]  Randomly choose the functionality $F$  
          \item[(c).]  \textbf{if} $F : x \ge y$ \textbf{then}  $E_{pk}(d) \gets E_{pk}(x-y)$\\                     
                   \textbf{if} $F : y \ge x+1$ \textbf{then}  $E_{pk}(d) \gets E_{pk}(y-x-1)$                   
          \item[(d).] $\delta \gets E_{pk}(d)$  
    \end{enumerate}
\STATE \textbf{for} $i=1$ to $m$ \textbf{do:}
\begin{enumerate}\itemsep=-1pt
        \item[(a).] $P_1$:
          \begin{itemize}\itemsep=-1pt
            \item $\tau_i \gets \delta \ast E_{pk}(r_i)$, where $r_i \in_R \mathbb{Z}_N$
            \item Send $\tau_i$ to $P_2$      
          \end{itemize} 
        \item[(b).] $P_2$:  
          \begin{itemize} \itemsep=-1pt 
                \item  $\tau'_i \gets D_{sk}(\tau_i)$
                \item  \textbf{if} $\tau'_i$ is even \textbf{then} $s_i \gets E_{pk}(0)$\\
                       \textbf{else} $s_i \gets E_{pk}(1)$                                                  
                \item Send $s_i$ to $P_1$
          \end{itemize}        
        \item[(c).] $P_1$:  
           \begin{itemize}\itemsep=-1pt
                \item  \textbf{if} $r_i$ is even \textbf{then}                     
                     $E_{pk}(d_i) \gets s_i$\\
                     \textbf{else} $E_{pk}(d_i) \gets E_{pk}(1)\ast s_i^{N-1}$
                \item $E_{pk}(d') \gets E_{pk}(d')\ast E_{pk}(d_i)^{2^{i-1}}$\\
                  \COMMENT{update $\delta$}
                \item $\Phi \gets \delta \ast E_{pk}(d_i)^{N-1}$ and $\delta \gets \Phi^{l}$ 
                \item \textbf{if} $i=m$ \textbf{then}\\
                  \begin{itemize}\itemsep=-1pt
                        \item $G \gets E_{pk}(d)\ast E_{pk}(d')^{N-1}$
                        \item $G' \gets G^r$, where $r \in_R \mathbb{Z}_N$ 
                        \item Send $G'$ to $P_2$  
                  \end{itemize}        
           \end{itemize}  
\end{enumerate}
\STATE $P_2$:
\begin{enumerate}\itemsep=-1pt
        \item[(a).] Receive $G'$ from $P_1$
        \item[(b).] \textbf{if} $D_{sk}(G') = 0$ \textbf{then}
                            $c'\gets 1$\\
                    \textbf{else} $c'\gets 0$
        \item[(c).] Send $E_{pk}(c')$ to $P_1$
\end{enumerate}
\STATE $P_1$:
\begin{enumerate}\itemsep=-1pt
        \item[(a).] Receive $E_{pk}(c')$ from $P_2$
        \item[(b).] \textbf{if} $F : x \ge y$ \textbf{then} $E_{pk}(c) \gets E_{pk}(c')$\\
                    \textbf{if} $F : y \ge x+1$ \textbf{then}  $E_{pk}(c) \gets E_{pk}(1)*E_{pk}(c')^{N-1}$ 
\end{enumerate}
\end{algorithmic}
\caption{\scp$(E_{pk}(x), E_{pk}(y)) \rightarrow E_{pk}(c)$ }
\label{alg:scp}
\end{algorithm}
\setlength{\textfloatsep}{0pt}
The overall steps involved in the proposed \scp~protocol are given in Algorithm 
\ref{alg:scp}. The basic idea of \scp~is 
for $P_1$ to randomly choose the functionality $F$ (by flipping a coin), where 
$F$ is either $x \ge y$ or $y \ge x+1$, and to obliviously execute $F$ with 
$P_2$. Briefly, depending on $F$, $P_1$ initially computes the encryption of difference 
between $x$ and $y$, say $d$. Then, $P_1$ and $P_2$ collaboratively decide the output based on whether 
$d$ lies in $[0,2^{m})$ or $[N-2^{m}, N)$. 
Since $F$ is randomly chosen and known only to $P_1$, the output of 
functionality $F$ remains oblivious to $P_2$. Before explaining the steps of \scp~in detail, we 
first discuss the basic ideas underlying our scheme which 
follow from Observations \ref{ob:1}, \ref{ob:2}, and \ref{ob:3}. 
\begin{observation}\label{ob:1}
For any given $x$ and $y$ such that $0 \le x,y < 2^m$, we know 
that $0 \le d < 2^m$ if $x \ge y$ and $N -2^{m} \le d < N$ otherwise, where $d = x-y$. Note 
that ``$N-y$'' is equivalent to ``$-y$'' under $\mathbb{Z}_N$. Then, 
we observe that $d - d'= 0$ only if $x \ge y$, where $d'$ denotes the integer corresponding to 
the $m$ least significant bits of $d$. On the other hand, if $x < y$, then we have 
$d - d' > 0$. 
\end{observation}
The above observation is clear from the fact that $d'$ always lies in $[0, 2^m)$. 
On one hand, when $x \ge y$, we have $d' =d$ since $d \in [0, 2^m)$. 
On the other hand, if $x < y$, we have $d\in [N-2^{m}, N)$; therefore, $d > d'$.
\begin{observation}\label{ob:2}
For any given $x$, let $x' = x+r \mod N$, where $r$ is a random number in  
$Z_N$ (denoted by $r\in_R \mathbb{Z}_N$). Here the relation between $x'$ and $r$ depends on whether $x + r \mod N$ leads 
to an overflow (i.e., $x + r$ is greater than $N$) or not. We observe 
that $x'$ is always greater than $r$ if there is no overflow. 
In the case of overflow, $x'$ is always less than $r$.
\end{observation}
The above observation is because of the fact that $x' = x+ r$ if there is no overflow. 
On the other hand, if there is an overflow, then $x' = x +r - N$.
\begin{observation}\label{ob:3}
For any given $x' = x+ r \mod N$, where $N$ is odd, the following property regarding the 
least significant bit of $x$ (denoted by $x_0$) always hold:
\[
x_0 = \left\{
\begin{array}{l l}
  \lambda_1\oplus \lambda_2 & \quad \text{if $r$ is even }\\
  1 - (\lambda_1\oplus \lambda_2) & \quad \text{otherwise}\\
\end{array} \right.
\]
\end{observation}
Here $\lambda_1$ denotes whether an overflow occurs or not, and $\lambda_2$ denotes 
whether $x'$ is odd or not. 
That is, $\lambda_1 = 1$ if $r > x'$ (i.e., overflow), and $0$ otherwise. Similarly, 
$\lambda_2=1$ if $x'$ is odd, and 0 otherwise. Observe that $1 - (\lambda_1\oplus \lambda_2)$ denotes the 
negation of bit $\lambda_1\oplus \lambda_2$. Also note 
that the RSA modulus $N$, which is a product of two 
large prime numbers, is always odd in the Paillier cryptosystem\cite{paillier-99}.

By utilizing the above observations, the proposed \scp~protocol aims to securely compute $E_{pk}(d-d')$ and 
check whether $d -d'=0$ or not. To start with, $P_1$ initially computes the 
multiplicative inverse of 2 under 
$\mathbb{Z}_N$ and assigns it to $l$. In addition, he/she sets $d'$ to 0. Then, $P_1$ 
chooses the functionality $F$ as either $x \ge y$ or $y \ge x+1$ 
randomly. Depending on $F$, $P_1$ computes the encryption of difference 
between $x$ and $y$ using 
homomorphic properties\footnote{In Paillier cryptosystem, ciphertext multiplications 
are followed by modulo $N^2$ operation so that 
the resulting ciphertext is still in $\mathbb{Z}_{N^2}$. However, 
to avoid cluttering the presentation, we simply omit 
the modulo operations.} as below:
\begin{itemize}
\item If $F: x \ge y$
\begin{eqnarray*}
E_{pk}(d) &=& E_{pk}(x)\ast E_{pk}(y)^{N-1} \\
         &=& E_{pk}(x-y)
\end{eqnarray*}
\item If $F: y \ge x+1$
\begin{eqnarray*}
E_{pk}(d) &=& E_{pk}(y)\ast E_{pk}(x+1)^{N-1} \\
         &=& E_{pk}(y-x-1)
\end{eqnarray*}
\item Observe that if $F: x\ge y$, then $d-d'=0$ only if $ x \ge y$. 
Similarly, if $F: y \ge x+1$, then $d-d'=0$ only 
if $y \ge x+1$.
\item Assign $E_{pk}(d)$ to $\delta$.
\end{itemize}
After this, $P_1$ and $P_2$ jointly compute $E_{pk}(d')$ in 
an iterative fashion. More specifically, at the end of  iteration $i$, $P_1$ knows 
the encryption of $i^{th}$ least significant bit as well as the encryption of 
integer corresponding to the $i$ least significant bits of $d$, for 
$1 \le i \le m$. Without loss of generality, let $d_i$ denote the $i^{th}$ least significant 
bit of $d$. Then, we have $d' = \sum_{i=1}^{m} d_{i}\ast 2^{i-1}$. In 
the first iteration, $P_1$ randomizes $\delta = E_{pk}(d)$ by 
computing $\tau_1 = \delta\ast E_{pk}(r_1)$ and sends it to $P_2$, where $r_1$ is a random 
number in $\mathbb{Z}_N$. Upon receiving $\tau_1$, $P_2$ decrypts it to get $\tau'_1 = 
D_{sk}(\tau_1)$ and checks its value. Note that 
$\tau'_1 = d + r_1 \bmod N$. Following from Observation \ref{ob:3}, 
if $\tau'_1$ is odd, $P_2$ computes $s_1 = E_{pk}(1)$, 
else he/she computes $s_1 = E_{pk}(0)$, and sends it to $P_1$. 
Observe that $s_1 = E_{pk}(\lambda_2)$. Also, to compute $\lambda_1$, we  need to perform secure comparison 
between $r_1$ and $\tau'_1$. However, in this paper, 
we assume that $\lambda_1$ is always zero (i.e., no overflow).
We emphasize that though we assume no overflow, $\tau'_i = d+ r_1 \mod N$ 
can still have overflow which depends on the actual 
values of $d$ and $r_1$. Nevertheless, in the later parts of this section, 
we show that for many practical applications, 
the above probabilistic assumption is very reasonable. 

Once $P_1$ receives $s_1$ from $P_2$, he/she computes $E_{pk}(d_1)$, encryption 
of the least significant bit of $d$, depending on 
whether $r_1$ is even or odd as below. 
\begin{itemize}
\item If $r_1$ is even, then $E_{pk}(d_1) = s_1 = E_{pk}(\lambda_2)$. 
\item Else $E_{pk}(d_1) = E_{pk}(1)\ast s_1^{N-1} \mod N^2 = E_{pk}(1- \lambda_2)$
\end{itemize}
Since $\lambda_1$ is assumed to be 0, following 
from Observation \ref{ob:3}, we have $\lambda_1 \oplus \lambda_2 = \lambda_2$. Also, 
note that ``$N-1$'' is equivalent to ``-1'' under $\mathbb{Z}_N$. Then, $P_1$ updates $E_{pk}(d')$ 
to $E_{pk}(d_1)$. 
After this, 
$P_1$ updates $\delta$ to $E_{pk}(\left \lfloor \frac{d}{2} \right \rfloor)$, encryption of 
quotient when $d$ is divided by 2, 
by performing following homomorphic additions: 
\begin{itemize}
\item $\Phi = \delta \ast E_{pk}(d_1)^{N-1} = E_{pk}(d - d_1)$
\item $\delta = \Phi^{l} = E_{pk}((d - d_1)\ast 2^{-1}) = E_{pk}(\left \lfloor \frac{d}{2} \right \rfloor)$
\end{itemize}
The main observation is that $ d-d_1$ is always a multiple 
of 2; therefore, $(d - d_1)\ast 2^{-1}$ always gives the correct quotient under $\mathbb{Z}_N$. 
The above process is continued iteratively such that in iteration $i$, $P_1$ knows 
$E_{pk}(d') = E_{pk}(\sum_{j=1}^{i} d_j \ast 2^{j-1})$ and updates 
$\delta$ accordingly, for $1 \leq i \leq m$.

In the last iteration, $P_1$ computes the encryption of difference between $d$ and $d'$ 
as $G = E_{pk}(d)* E_{pk}(d')^{N-1} = E_{pk}(d-d')$. Then, he/she randomizes $G$ 
by computing $G' = G^r$ and 
sends it to $P_2$, where $r$ is a random number in $\mathbb{Z}_N$. After this, $P_2$ decrypts $G'$ 
and sets $c'=1$ if $D_{sk}(G') = 0$, and $c'=0$ otherwise. Also, $P_2$ sends $E_{pk}(c')$ to $P_1$. 
Finally, depending on $F$, $P_1$ computes the output $E_{pk}(c)$ as below:
\begin{itemize}
\item If $F: x \ge y$, then set $E_{pk}(c)$ to $E_{pk}(c')$.
\item Otherwise, compute the negation of $E_{pk}(c')$ and  
assign it to $E_{pk}(c)$. That is, $E_{pk}(c) = E_{pk}(1)*E_{pk}(c')^{N-1} = E_{pk}(1-c')$.
\end{itemize}
\begin{example}
Suppose $x = 1, y = 5,$ and $m=3$. Let us assume that $P_1$ holds 
$(E_{pk}(1), E_{pk}(5))$.  Under this case, we show various intermediate results 
during the execution of the proposed \scp~protocol. Without loss 
of generality, we assume that $P_1$ chooses the functionality $F: y \ge x+1$. Initially, 
$P_1$ computes $E_{pk}(d)=E_{pk}(3)$ and sets 
it to $\delta$. Since $m=3$, \scp~computes $E_{pk}(d')$ in three iterations. 
For simplicity, we assume 
that $r_i$'s are even and there is no overflow. Note 
that, however, $r_i$ is different in each iteration.
\begin{eqnarray*}
\textbf{Iteration 1:} & & \\
\tau'_1 & = & 3 + r_1 \mod N = an~odd~integer\\
E_{pk}(d_1) & = & s_1 = E_{pk}(1)\\
E_{pk}(d') & = & E_{pk}(d_1) = E_{pk}(1)\\
\delta & = & E_{pk}((3 -  1)\ast 2^{-1}) =  E_{pk}(1)\\
\textbf{Iteration 2:} & & \\
\tau'_2 & = & 1 + r_2 \mod N = an~odd~integer\\
E_{pk}(d_2) & = & s_2 = E_{pk}(1)\\
E_{pk}(d') & = & E_{pk}(d')\ast E_{pk}(d_2)^2 = E_{pk}(3)\\
\delta & = & E_{pk}((1 -  1)\ast 2^{-1}) =  E_{pk}(0)\\
\textbf{Iteration 3:} & & \\
\tau'_3 & = &  r_3 \mod N = an~even~integer\\
E_{pk}(d_3) & = & s_3 = E_{pk}(0)\\
E_{pk}(d') & = & E_{pk}(d')\ast E_{pk}(d_3)^4 = E_{pk}(3)\\
\delta & = & E_{pk}(0)\\
\end{eqnarray*}
At the end of the 3rd iteration, $P_1$ has $E_{pk}(d') = E_{pk}(3) = E_{pk}(d)$. 
After this, $P_1$ computes $G' = E_{pk}( r*(d-d')) = E_{pk}(0)$ and sends it 
to $P_2$. Upon receiving, $P_2$ decrypts $G'$ to get 0, sets $c'$ to 1, and sends 
$E_{pk}(c')$ to $P_1$. Finally, $P_1$ computes $E_{pk}(c) = E_{pk}(1 -c') = E_{pk}(0)$. It 
is clear that, since $x <y$, we have $c=0$.
\hfill $\Box$
\end{example}

\subsection{Correctness Analysis}
In this sub-section, we theoretically prove that 
our \scp~scheme generates the correct result with very high 
probability. First, we emphasize that the correctness of \scp~depends 
on how accurately can $P_1$ and $P_2$ compute $E_{pk}(d')$. Since 
$d' = \sum_{i=1}^m d_i*2^{m-1}$ is computed in an iterative fashion, this further implies 
that the correctness depends on the accuracy of the least $m$ significant 
bits computed from $d$. 

In each iteration, $r_i$ 
can take any value in $\mathbb{Z}_N$. We observe that if $r_i\in [N - 2^m, N)$, 
only then the corresponding computed encrypted bit of $d$, i.e., $E_{pk}(d_i)$ 
can be wrong (due to overflow). That is, the number of possible values of $r_i$ 
that can give rise to error are $2^{m}$. 
Since we have $N$ number of possible values for $r_i$, the probability for producing 
wrong bit is $\frac{2^{m}}{N} \approx \frac{1}{2^{K-m}}$, where $K$ is 
the encryption key size in bits. Therefore, the 
probability for computing the encryption of $d_i$ correctly 
is approximately $1 - \frac{1}{2^{K-m}}$. This 
probability remains the same for all the bits since $r_i$ is chosen independently in each iteration. 
Hence, the probability for \scp~to compute the correct 
value of $E_{pk}(d')$ is given by: 
$$\left (1 - \frac{1}{2^{K-m}} \right )^m \approx  {e^{-\frac{m}{2^{K-m}}}}$$

In general, for many real-world applications, $m$ can be 
at most 100 (since $0 \le x,y < 2^{100}$ is sufficiently large enough to suit most applications). 
Therefore, for 1024-bit key size, 
the probability for \scp~to produce the correct output is approximately $e^{-\frac{100}{2^{924}}} 
\approx 1$. Hence, for practical domain values of $x$ and $y$, 
with a probability of almost 1, the \scp~protocol gives 
the correct output $E_{pk}(d')$. We 
emphasize that even in the extreme case, such as $m=950$, 
the probability for \scp~to produce 
correct $E_{pk}(d')$ is $e^{-\frac{950}{2^{74}}} \approx 1$. 

Additionally, following from Observation \ref{ob:1}, the value of $d-d'$ is 
equal to 0 iff the corresponding functionality under $F$ is true. In practice, as 
mentioned above, we have $K > m$. When $F$ is true, the property 
$0 \le d,d' < 2^m$ holds and the integer corresponding to the $m$ least significant bits 
of $d$ is always equivalent to $d'$. Therefore, the decryption of $G' = E_{pk}(r\ast(d-d'))$ by $P_2$ 
will result in 0 iff $F$ is true. In particular, when $F:y \ge x+1$ , the negation operation by $P_1$ makes 
sure that the final output is equal to $E_{pk}(c)$. 
Hence, based on the above discussions, it is clear  
that the proposed \scp~scheme produces correct result with very high probability. 

\subsection{Security Analysis}\label{security-scp}
\subsubsection{Proof of Security under the Semi-honest Model}
The security goal of \scp~is to prevent $P_1$  and $P_2$ 
from knowing $x$ and $y$. In addition, the comparison result should be protected   
from both $P_1$ and $P_2$.  
Informally speaking, since $d$ is the only value related to $x$ and $y$,
either $d$ or part of $d$ is always hidden by a random number; therefore, $P_2$
does not know anything about $d$. As a result, $P_2$ knows nothing about 
$x$ and $y$. On the other hand, since $P_1$ does not have the decryption key
and the comparison result is encrypted, $P_1$ does not know $x\ge y$ or $y \ge x+1$. 
Moreover, because $P_1$ randomly selects which functionality between $x\ge y$
and $y\ge x+1$ to compute, $P_2$ does not know the comparison result either. 
However, we may ask why to randomly select a functionality. If we do not, $P_2$ will know,
for example, the first value is bigger than the second value. This seemingly useless information
can actually allow $P_2$ to learn data access patterns which in turn breaks 
the semantic security of the underlying encryption scheme. 
Next we provide a formal proof of security for \scp~under the semi-honest model.

As stated in Section \ref{subsec:secure-def}, to prove the security of the proposed
protocol under the semi-honest setting, we adopt the well-known security 
definitions and techniques in the literature 
of secure multiparty computation.
To formally prove \scp~is secure\cite{goldreich-general}, we need to show that the simulated execution image of 
\scp~is computationally indistinguishable from the actual execution 
image of \scp. An execution image generally includes
the messages exchanged and the information computed from these messages.
Therefore, according to Algorithm \ref{alg:scp}, the execution image of 
$P_2$ can be denoted by $\Pi_{P_2}$, where
$$
\Pi_{P_2} = \{\langle E_{pk}(\delta+r_i), \delta + 
r_i~\textrm{mod}~N \rangle, \langle G', b \rangle|~\textrm{for}~ 1 \le i \le m\} 
$$
Note that $\delta + r_i \mod N$ is derived from $E_{pk}(\delta + r_i)$, where the modulo 
operator is implicit in the decryption function. $P_2$ receives $G'$ at the last iteration
and $b$ denotes the decryption result of $G'$.
Let the simulated image of $P_2$ be $\Pi^S_{P_2}$, where
$$
\Pi^S_{P_2} = \{\langle s_i^1, s_i^2\rangle, \langle s, b'\rangle|~\textrm{for}~ 1 \le i \le m\} 
$$
Both $s_i^1$ and $s$ are randomly generated from 
$\mathbb{Z}_{N^2}$, and  
 $s_i^2$ is randomly generated from  $\mathbb{Z}_{N}$. 
Since $E_{pk}$ is a semantically 
secure encryption scheme with resulting ciphertext size less than $N^2$, 
$E_{pk}(\delta + r_i)$ and $G'$ are computationally indistinguishable from $s_i^1$ and $s$, respectively. Also,
as $r_i$ is randomly generated,  $\delta + r_i \mod N$ 
is computationally indistinguishable from $s_i^2$. Furthermore, 
because the functionality is randomly chosen by $P_1$ (at step 1(b) of Algorithm 
\ref{alg:scp}), $b$ is either 0 or 1 with equal probability.  Thus,
 $b$ is computationally indistinguishable from $b'$. Combining all these results 
together, we can conclude that 
$\Pi_{P_2}$ is computationally indistinguishable from $\Pi^S_{P_2}$.
This implies that during the execution of \scp, $P_2$ does not learn anything about $x$
and $y$. Intuitively speaking, the information $P_2$ has during an execution of \scp~is
either random or pseudo-random, so this information 
does not disclose anything regarding $x$ and $y$.

Similarly, the execution image of 
$P_1$ can be denoted by $\Pi_{P_1}$, where
$$
\Pi_{P_1} = \{s_i, E_{pk}(c')|~\textrm{for}~ 1 \le i \le m\} 
$$
Let the simulated image of $P_1$ be $\Pi^S_{P_1}$, where
$$
\Pi^S_{P_1} = \{s_i', s|~\textrm{for}~ 1 \le i \le m\} 
$$
Both $s_i$ and $s$ are randomly generated from 
$\mathbb{Z}_{N^2}$. 
Since $E_{pk}$ is a semantically 
secure encryption scheme with resulting ciphertext size less than $N^2$, 
$s_i$ and $E_{pk}(c')$ are computationally indistinguishable from $s_i'$ and $s$, respectively. 
Therefore, $\Pi_{P_1}$ is computationally indistinguishable from $\Pi^S_{P_1}$.
This implies that $P_1$ does not learn anything about the comparison result.
Combining with previous analysis, we can say \scp~is secure under the semi-honest model.
%
%

\subsubsection{Security against Malicious Adversary}
After proving that \scp~is secure under the semi-honest model, 
the next step is to extend it to a secure protocol against malicious adversaries. Under 
the malicious model, an adversary (i.e., either $P_1$ or $P_2$) can arbitrarily deviate 
from the protocol to gain some advantage (e.g., learning additional information about inputs) 
over the other party. The deviations include, as an example, for $P_1$ (acting as a malicious adversary) 
to instantiate the \scp~protocol with modified inputs $(E_{pk}(x'), E_{pk}(y'))$ and to 
abort the protocol after gaining partial information. 
However, in \scp, it is worth pointing out that neither $P_1$ nor $P_2$ knows the comparison 
result. In addition, all the intermediate results are either random or pseudo-random values. 
Thus, even when an adversary modifies the intermediate computations he/she cannot gain 
any information regarding $x, y,$ and $c$. Nevertheless, as mentioned above, 
the adversary can change the intermediate data or perform computations incorrectly 
before sending them to the honest party which may eventually result in the wrong output. 
Therefore, we need to ensure that all the computations performed and messages 
sent by each party are correct. We now discuss two 
different approaches from the literature to extend 
the \scp~protocol and make it secure under the malicious model. 

The standard way of preventing the malicious party from 
misbehaving is to let the honest party validate the other party's work using    
zero-knowledge proofs\cite{camenisch-1999}. First of all, we stress that input 
modification in any secure protocol cannot be prevented\cite{goldreich-encryption}; therefore,  
we proceed as follows. On one hand, if $P_1$ is a malicious adversary and 
the input of \scp~is generated 
as a part of an intermediate step, then the honest party (i.e., $P_2$) can validate it correctness using zero knowledge 
proofs. On the other hand, where $E_{pk}(x)$ and $E_{pk}(y)$ are not part of an intermediate step, 
we assume that the input is committed (e.g., explicitly certified by the data owner). Under this case, 
the honest party can validate the intermediate computations of $P_1$ based on the committed input. 
Also, we assume that there exist no collusion between $P_1$ and $P_2$ (i.e., at most one party is 
malicious). Note that such 
an assumption is necessary to construct secure protocols under the malicious model. 

Recently, Nikolaenko et al.\cite{nikolaenko-2013} discussed a mechanism for the honest party 
to validate the data sent by the adversary under (asymmetric) two-party setting. Their approach utilizes Pedersen 
commitments\cite{pedersen-1991} along with the zero-knowledge proofs to prove modular arithmetic relations between 
the committed values. However, checking the validity of computations at each step of \scp~can 
significantly increase the overall cost.

An alternative approach, as proposed in \cite{huang-2012}, is to instantiate two 
independent executions of the \scp~protocol by swapping the roles 
of the two parties in each execution. At the end of the individual executions, each party receives 
the output in encrypted form. This is followed by an equality test on their outputs. 
More specifically, suppose $E_{pk_1}(c_1)$ and $E_{pk_2}(c_2)$ be the outputs received 
by $P_1$ and $P_2$ respectively, where $pk_1$ and $pk_2$ are their respective public keys. 
A simple equality test on $c_1$ and $c_2$, which produces an output value of 1 if $c_1 = c_2$ and a random number otherwise,  
is sufficient to catch the malicious adversary. That is, the malicious party, which will be 
caught in the case of cheating, acts as a covert adversary\cite{aumann-2010}. Under 
the covert adversary model, the parties can shift the verification step until the end and 
then directly compare the final outputs. We emphasize 
that the equality test based on the additive homomorphic encryption properties  which 
was used in \cite{huang-2012} is not applicable 
to our problem. This is because, the outputs in our case are in encrypted format and the corresponding 
ciphertexts (resulted 
from the two executions) are under two 
different public key domains. Nevertheless, $P_1$ and $P_2$ can perform 
the equality test by constructing a garbled-circuit based on the similar steps as mentioned in the 
\scg~protocol.  

\subsection{Performance Comparison of \scp~with Existing Work}\label{subsec:sc-exp}
In this sub-section, we empirically compare the computation costs 
of \scp~with those of \scg. As discussed in Section \ref{subsec:sc-disc}, 
the \scg~protocol is based on the Yao's garbled-circuit technique.   
Besides \scg, another well-known solution to the secure comparison 
of Paillier encrypted integers was proposed by Blake et al.\cite{blake-2009}. However, as mentioned earlier, their 
protocol requires the encryptions of bits rather 
than pure integers as the inputs. Nevertheless, 
one could combine their protocol with the existing secure bit-decomposition (SBD) methods 
to solve the SC problem. Recently, Samanthula et al.\cite{bksam-cloud13}  
proposed a new SBD method and combined it with \cite{blake-2009} to solve 
the SC problem. We denote  
such a construction by \scb. To the best of our knowledge, \scb~is the most efficient 
custom-designed method (under Paillier cryptosystem) to perform secure comparison over encrypted integers. 

To better understand the efficiency gains of \scp, we need to 
compare its computation costs with both \scg~and \scb. For this purpose, we 
implemented all the three protocols in C using Paillier's scheme\cite{paillier-99} and 
conducted experiments on a 
Intel\textregistered~Xeon\textregistered~Six-Core\texttrademark~3.07GHz 
PC with 12GB memory running Ubuntu 10.04 LTS. In particular to \scg, 
we constructed and evaluated the circuit using FastGC\cite{huang-2011} 
framework on the same machine. Since \scb~is secure 
under the semi-honest model, in our implementation we assume 
the semi-honest setting for a fair comparison among the three protocols. 
That is, we did not implement the extensions to \scp~that are secure under the malicious setting. 
\begin{figure}[t]
\centering
\includegraphics[width=.5\textwidth]{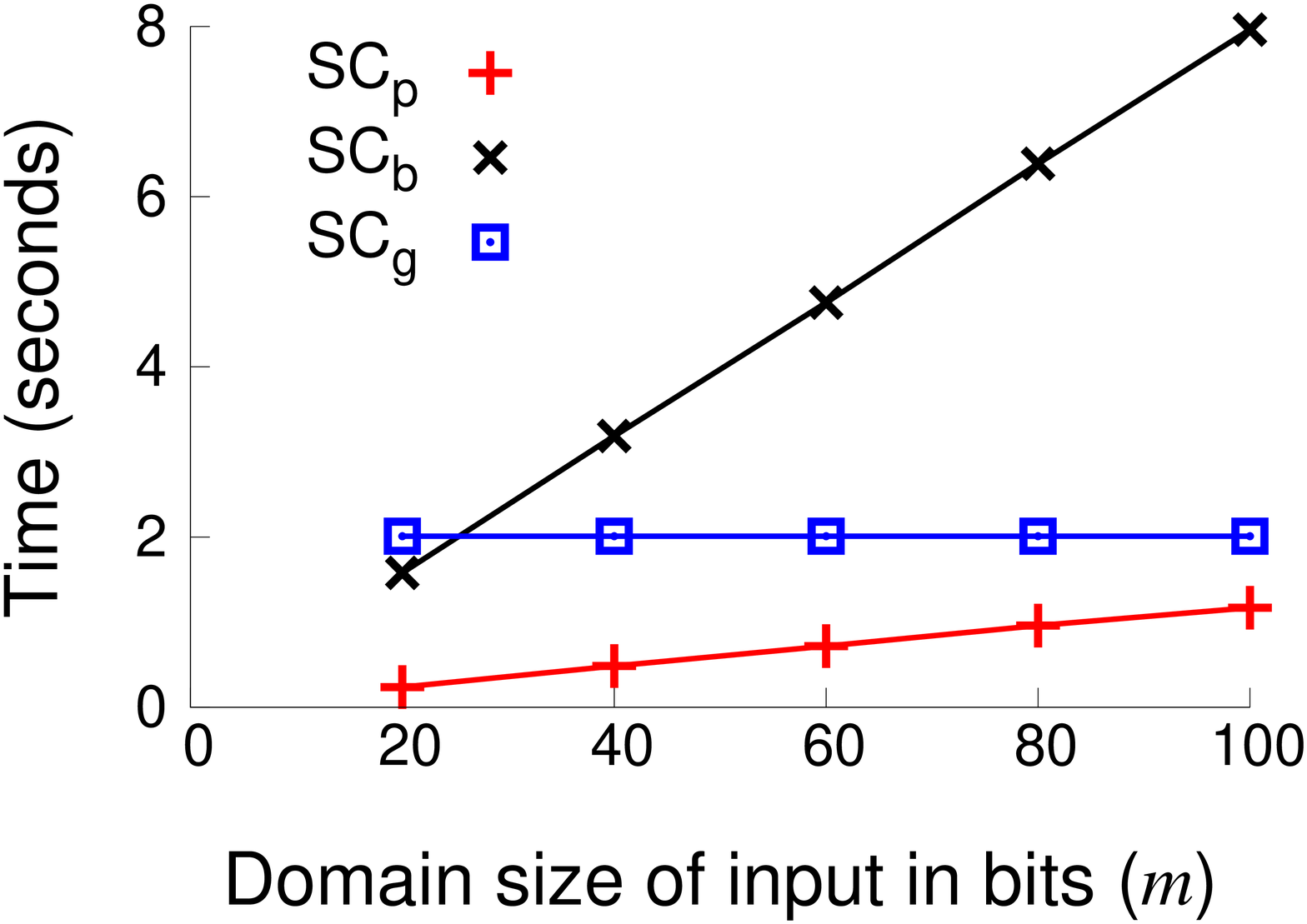}
\caption{Comparison of computation costs of \scp~with those of \scg~and \scb~for $K=1024$ bits and varying $m$}
\label{fig:comp-sc-1024}
\end{figure}

For encryption key size $K=1024$ bits (a commonly accepted key size which 
also offers the same security guarantee as in FastGC\cite{huang-2011}), the comparison results are as shown in 
Figure \ref{fig:comp-sc-1024}. Following from Figure \ref{fig:comp-sc-1024}, 
it is clear that the computation costs of both \scp~and \scb~grow linearly with the 
domain size $m$ (in bits) whereas the computation cost of \scg~remains constant at 2.01 seconds. This 
is because, the \scg~protocol uses random shares as input instead of 
encryptions of $x$ and $y$. On one hand, the computation costs 
of \scb~varies from 1.58 to 7.96 seconds when $m$ is changed from 20 to 100. 
On the other hand, the computation costs of \scp~increases from 0.23 to 1.16 seconds 
when $m$ is changed from 20 to 100. It is evident that \scp~outperforms both the protocols irrespective 
of the value of $m$. Also, for all values of $m$, we observe that \scp~is at least 
6 times more efficient than \scb. In addition, when $m=20$, 
our \scp~protocol is around  
8 times more efficient than the circuit-based \scg~protocol. 

From a privacy perspective, it is important to note that \scp~and \scg~guarantee 
the same level of security by not revealing the input values as well as the comparison result 
to $P_1$ and $P_2$. Although \scb~leaks the comparison result to at least 
one of the participating parties, as mentioned in Section 
\ref{subsec:sc-disc}, it can be easily modified to our setting at the expense of additional cost. 
Since \scp~provides similar security guarantee, but more efficient than \scg~and \scb, 
we claim that \scp~can be used as a building block 
in larger privacy-preserving applications, such as secure clustering, to boost the overall 
throughput by a significant factor. 
Furthermore, we emphasize that our \scp~scheme is more reliable than 
\scb~in terms of round complexity. More specifically, \scp~require 
$m$+1 number of communication rounds whereas \scb~require 
2$m$+1 number of communication rounds between $P_1$ and $P_2$. On the other hand, \scg~requires 
constant number of rounds. Nevertheless, we would 
like to point out that the 
round complexity of our \scp~scheme can be reduced to (small) constant 
number of rounds by using the Carry-Lookahead Adder\cite{comparch} with similar 
computation costs. Since the constant-round \scp~protocol is much more complex to present and 
due to space limitations, in this paper, we presented 
the \scp~protocol whose round complexity is bounded by $O(m)$. Also, 
we emphasize that the efficiency of \scp~(and other SC protocols) can be improved 
further by using alternative HEnc$^+$ systems (e.g., \cite{damgaard-2007}) which 
provide faster encryption than a Paillier encryption. 

\section{The Proposed PPRQ Protocols}\label{sec:proposed}
In this section, we propose two novel PPRQ protocols over encrypted data 
in the cloud computing environment. Our protocols utilize the above-mentioned \scp~scheme 
and secure multiplication (SMP) as the building blocks. 
In addition, we analyze the security guarantees and complexities of 
the proposed protocols in detail. The two protocols 
act as a trade-off between efficiency and flexibility. In particular, 
our second protocol incurs negligible computation cost on the end-user. 

Both protocols consider two cloud 
service providers denoted by $C_1$ (referred to as primary cloud) and $C_2$ (referred to as 
secondary cloud) which together form a federated cloud \cite{buyya-2010}. 
As justified in Section \ref{subsec:secure-def}, for the rest of this paper, we 
assume that the probability of collusion between $C_1$ and 
$C_2$ is negligible (which is reasonable in practice). We emphasize that 
such an assumption has been commonly used in the related 
problem domains (e.g., \cite{twinclouds-2011}). The main intuition 
behind this assumption is as follows. Suppose the two servers can be 
implemented by two cloud service providers, such as Google and Amazon. Then it is hard to 
imagine why Google and Amazon want to collude to damage their reputation which 
could cost billions to repair.

Under the above cloud setting, Alice initially generates a 
Paillier public-secret key pair $(pk, sk)$ and sends the secret key $sk$ 
to $C_2$ through a secure channel whereas 
$pk$ is treated as public information. 
Additionally, we explicitly make the 
following practical assumptions in our problem setting:
\begin{itemize}
\item Alice encrypts her database $D$ attribute-wise 
using her public key $pk$. More specifically, she computes $T_{i,j} = E_{pk}(t_{i,j})$, 
where $t_{i,j}$ denotes the $j^{th}$ attribute value of data record $t_i$, for $1 \le i \le n$ and 
$1 \le j \le w$. After this, she outsources the encrypted database $T$ to $C_1$. It 
is important to note that the cost (both computation and communication) 
incurred on Alice during this step is a one-time cost. In the proposed protocols, 
after outsourcing $T$ to $C_1$, Alice can remain offline since the entire 
query processing task is performed by $C_1$ and $C_2$.
\item The attribute values lie in $[0, 2^m)$, where $m$ is the 
domain size of the attributes (in bits). In general, $m$ may vary for each attribute. However, for security 
reasons, we assume that $m$ is the same for all attributes. One way of selecting $m$ is to take 
the maximum out of all attribute domain sizes. For the rest of this paper, 
we assume $m$ is public\footnote{For a better security, the data owner Alice can mask $m$ by adding 
a small random number $m'$ (where both $m$ and $m'$ are known only to Alice) to it. Under this case, the value of 
$m+m'$ can be treated as public information.}. 
\item We assume that the number of data records (i.e., $n$) and attributes (i.e., $w$) can be revealed to the clouds. We 
emphasize that Alice can include some dummy records to $D$ (to hide $n$) and dummy attributes 
to each record (to hide $w$). However, for simplicity, we assume that 
Alice does not add any dummy records and attributes to $D$. The values of $n$ and $w$ are treated as public.
\item All parties are assumed to be semi-honest and there is no 
collusion between different parties. However, we stress that by combining the malicious \scp~protocol with 
zero-knowledge proofs, we can easily extend our protocols to secure protocols under the malicious model. 
Also, we assume that there exist secure communication channels 
between each pair of parties involved in our protocols. Note that the existing secure mechanisms, 
such as SSL, can be utilized for this purpose.
\item We assume that the set of authorized users (decided 
by Alice) who can access $D$ is known to $C_1$ and $C_2$. This is a practical assumption 
as it will also be useful for them to verify users' identity during authentication\cite{chow-2009}.
\end{itemize}
We emphasize that the above assumptions are commonly made in the literature of 
related problem domains, and we do not make any abnormal assumptions.
\subsection{Protocol 1}
In the proposed first protocol, referred to as PPRQ$_1$, we assume that each authorized 
user generates a public-secret key pair. In particular, we 
denote Bob's public-secret key pair by $(pk_b, sk_b)$. 

After outsourcing the attribute-wise encrypted database of $D$ (i.e., $T$) by Alice to $C_1$, 
if at some future time, suppose Bob wants to perform a range query on the encrypted 
data in the cloud. Let $k$ be the attribute index upon which he wants to filter 
the records.  During the query request step, he first computes the 
additive random shares of lower and upper bound values in his query. That is, he 
computes 
random shares $\{\alpha_1, \alpha_2\}$ and $\{\beta_1, \beta_2\}$  such that 
$\alpha = \alpha_1 + \alpha_2 \bmod N$ and $\beta = \beta_1 + \beta_2 \bmod N$, where $\alpha$ and 
$\beta$ are the lower and upper bound values of his range query. 
Note that $0 \le \alpha, \beta < 2^m$. 
The goal here is for Bob to securely 
retrieve the data record $t_i$ only if $\alpha \leq t_{i,k} \leq \beta$, for $1 \leq i \leq n$. 
We emphasize that $\alpha$ and $\beta$ are 
private information of Bob; therefore, they should not be revealed to Alice, $C_1$ 
and $C_2$. 

\begin{algorithm}[!thbp]
\begin{algorithmic}[1]
\REQUIRE $sk$ is known only to Alice and $C_2$; $sk_b$ is known only to Bob; whereas $pk$ and 
$pk_b$ are public; $\pi$ is known only to $C_1$; $Q = \{k, \alpha, \beta\}$ is private to Bob\\
\COMMENT{\textbf{Step 1 - Query Request}}
\STATE Bob:
\begin{enumerate}\itemsep=-1pt
    \item[(a).] $\alpha_1 + \alpha_2 \bmod N \gets \alpha$ and $\beta_1 + \beta_2 \bmod N \gets \beta$
    \item[(b).]  Send $\{k, \alpha_1, \beta_1\}$ to $C_1$ and $\{\alpha_2, \beta_2\}$ to $C_2$
\end{enumerate}
\COMMENT{\textbf{Steps 2 to 5 - Data Processing}}
\STATE $C_2$ sends $\{E_{pk}(\alpha_2), E_{pk}(\beta_2)\}$ to $C_1$
\STATE $C_1$: 
\begin{enumerate}\itemsep=-1pt 
       \item[(a).] $E_{pk}(\alpha) \gets E_{pk}(\alpha_1)*E_{pk}(\alpha_2)$ and $E_{pk}(\beta) \gets E_{pk}(\beta_1)*E_{pk}(\beta_2)$
\end{enumerate}
\STATE $C_1$ and $C_2$, \textbf{for} $1 \leq i \leq n$ \textbf{do:}
\begin{enumerate}\itemsep=-1pt    
    \item[(a).]  $L_i \gets$~\scp$(T_{i,k}, E_{pk}(\alpha))$, here only $C_1$ receives $L_i$
    \item[(b).]  $M_i \gets$~\scp$(E_{pk}(\beta), T_{i,k})$, here only $C_1$ receives $M_i$
    \item[(c).]  $O_i \gets$~SMP$(L_i, M_i)$, here $O_i$ is known only to $C_1$
    \item[(d).]  $T'_{i,j} \gets$~SMP$(T_{i,j}, O_i)$, for $1 \le j \le w$ 
\end{enumerate}
\STATE $C_1$:
\begin{enumerate}\itemsep=-1pt    
    \item[(a).] \textbf{for} $1 \leq i \leq n$ and $1 \le j \le w$~\textbf{do:}
              \begin{itemize}\itemsep=-1pt
                \item $U_{i,j} \gets T'_{i,j}\ast E_{pk}(r_{i,j})$, where $r_{i,j} \in_R \mathbb{Z}_N$                 
                \item $V_{i,j} \gets E_{pk_b}(r_{i,j})$ 
              \end{itemize}  
    \item[(b).] Row-wise permutation: $X \gets \pi(U)$ and  $Y \gets \pi(V)$
    \item[(c).] $Z \gets \pi(O)$
    \item[(d).] Send $X, Y$ and $Z$ to $C_2$
\end{enumerate}
\COMMENT{\textbf{Step 6  - Query Response}}
\STATE $C_2$, \textbf{for} $1 \leq i \leq n$ \textbf{do}:
\begin{enumerate}\itemsep=-1pt
    \item[(a).]  \textbf{if} $D_{sk}(Z_i)=0$ \textbf{then:}
       \begin{itemize}\itemsep=2pt
         \item $x_{i,j} \gets D_{sk}(X_{i,j})$, for $1 \le j \le w$
         \item Send $(x_i, Y_i)$ to Bob
       \end{itemize}  
       \textbf{else}  Ignore $X_i$ and $Y_i$    
\end{enumerate}
\COMMENT{\textbf{Step 7 - Data Decryption}}
\STATE Bob: 
\begin{enumerate}\itemsep=1pt
    \item[(a).] $S\gets \emptyset$
    \item[(b).] \textbf{foreach} entry $(x_i, Y_i)$ received from $C_2$ \textbf{do}:
      \begin{itemize} \itemsep=1pt       
         \item $\gamma_{i,j} \gets D_{sk_b}(Y_{i,j}),$ for $1 \le j \le w$
         \item $t'_{j} \gets x_{i,j} - \gamma_{i,j} \bmod N$, for $1 \le j \le w$           
         \item $S \gets S \cup t'$  
      \end{itemize}
\end{enumerate}
\end{algorithmic}
\caption{$\textrm{PPRQ}_1(T, Q) \rightarrow S$}
\label{alg:pprq1}
\end{algorithm}
The overall steps involved in the proposed PPRQ$_1$ protocol 
are shown in Algorithm \ref{alg:pprq1}.  To start with, Bob initially sends $\{k, \alpha_1, \beta_1\}$ and $\{\alpha_2, \beta_2\}$ 
to $C_1$ and $C_2$, respectively. 
Upon receiving $\{\alpha_2, \beta_2\}$ from Bob\footnote{Note that if Bob is not an authorized user (which is usually 
decided by Alice), then $C_2$ simply dumps 
the query request of Bob.}, $C_2$ computes $\{E_{pk}(\alpha_2), E_{pk}(\beta_2)\}$ and sends it to $C_1$. 
Then, $C_1$ computes the encrypted values of $\alpha$ and $\beta$ locally using additive homomorphic 
properties. That is, $C_1$ computes $E_{pk}(\alpha)$ as $E_{pk}(\alpha_1)*E_{pk}(\alpha_2)$ and 
$E_{pk}(\beta)$ as $E_{pk}(\beta_1)*E_{pk}(\beta_2)$. After this, $C_1$  and $C_2$ 
jointly involve in the following set of operations, for 
$1 \le i \le n$: 
\begin{itemize}\itemsep=0pt
\item Securely compare $T_{i,k}$, i.e., the encryption of $k^{th}$ attribute 
value of data record $t_i$ in $D$, with $E_{pk}(\alpha)$ and $E_{pk}(\beta)$ using the 
\scp~protocol (in parallel). Without loss of generality, suppose $L_i =~$\scp$(T_{i,k}, E_{pk}(\alpha))$ and 
$M_i =~$\scp$(E_{pk}(\beta), T_{i,k})$. At the end of this step, the outputs $L_i$ and $M_i$, which are in encrypted 
format, are known only to $C_1$. 
\item Securely multiply $L_i$ and $M_i$ using the secure 
multiplication (SMP) protocol. The SMP protocol is one of the basic building blocks in the field of 
secure multiparty computation \cite{goldreich-general}. Briefly, given a party $P_1$ holding 
$(E_{pk}(a), E_{pk}(b))$ and a party $P_2$ with $sk$, the SMP protocol returns $E_{pk}(a*b)$ to $P_1$. During 
this process, no information regarding $a$ and $b$ is revealed to $P_1$ and $P_2$. An efficient  
implementation of SMP is given in the Appendix. Let $O_i$ denote the output 
of SMP$(L_i, M_i)$. The observation here is $O_i= E_{pk}(1)$ only if $L_i = M_i = E_{pk}(1)$. 
This further implies that $t_{i,k} \ge \alpha$ and $\beta \ge t_{i,k}$. Otherwise, $O_i = E_{pk}(0)$. The 
output $O_i$ is known only to $C_1$. Since $O_i$ is an encrypted value, neither $C_1$ nor $C_2$ know 
whether the corresponding record $t_i$ matches the query condition $\alpha \le t_{i,k} \le \beta $.
\item Generate a dataset $T'$ such that $T'_{i,j} =$ SMP$(T_{i,j}, O_i)$, for $1 \le j \le w$. We emphasize 
that $T'_{i,j}= T_{i,j}$ iff $O_i$ is an encryption of 1. That is, if index $i$ satisfies 
the property $\alpha \le t_{i,k} \le \beta$, then $T'_i=T_i$. Otherwise, all 
the entries in $T'_i$ are encryptions of 0's. At the end, the output $T'$ is known only to $C_1$. 
\end{itemize}
After this, $C_1$ locally involves in the following set of operations, for $1 \le i \le n$ and 
$1\le j \le w$:
\begin{itemize}\itemsep=0pt
\item Randomize $T'_{i,j}$ using additive homomorphic property to 
get $U_{i,j} =  T'_{i,j} \ast E_{pk}(r_{i,j})$, 
where $r_{i,j}$ is a random number in $\mathbb{Z}_N$. Also, encrypt the random 
number $r_{i,j}$ using Bob's public key $pk_b$ to get $V_{i,j} = E_{pk_b}(r_{i,j})$. 
\item Perform a row-wise permutation on $U$ and $V$ to get $X = \pi(U)$ and 
$Y = \pi(V)$. Here $\pi$ is a random permutation function 
known only to $C_1$. Also, $C_1$ randomly permutes the vector $O$, i.e. he/she  
computes $Z = \pi(O)$. Then, $C_1$ sends $X, Y,$ and $Z$ to $C_2$. 
\end{itemize}
Upon receiving, $C_2$ filters the entries of $X$ and $Y$ using $Z$ as follows. 
We observe that if $D_{sk}(Z_i) = 1$, i.e., $O_i = L_i = M_i= E_{pk}(1)$, then the $k^{th}$ column value of $t_{\pi^{-1}(i)}$ satisfies the 
input range query condition, for $1 \le i \le n$. This is because, when $L_i =E_{pk}(1)$, we have 
$t_{\pi^{-1}(i),k} \geq \alpha$. On the other hand, when $M_i = E_{pk}(1)$, 
we have $t_{\pi^{-1}(i),k} \leq \beta$. Therefore, 
when $O_i = L_i = M_i= E_{pk}(1)$, the desired condition $\alpha \leq t_{\pi^{-1}(i),k} \leq \beta$ always holds.
Hence, under this case, $C_2$ decrypts $X_i$ attribute-wise to get $x_{i,j} = D_{sk}(X_{i,j})$, for 
$1 \le j \le w$, and sends the entry $(x_i, Y_i)$ to 
Bob. Observe that $x_{i,j}$ is a random number in $Z_{N}$. On the other hand, if $D_{sk}(Z_i) = 0$, we have 
$L_i= E_{pk}(0)$ or $M_i= E_{pk}(0)$; therefore, the corresponding $k^{th}$ column value does not 
lie in $(\alpha, \beta)$. Hence, under this case, $C_2$ simply ignores $X _i$ and $Y_i$. Note that since $Z$ is a randomly 
permuted vector of $O$ and 
as $\pi$ is known only to $C_1$, $C_2$ cannot trace back which data record in $D$ corresponds to $Z_i$. 

After receiving the entries (if there exist any) from $C_2$, Bob 
initially sets the output set $S$ to $\emptyset$. Then, he 
proceeds as follows for each received entry $(x_i, Y_i)$ and $1 \le j \le w$:
\begin{itemize}\itemsep=0pt
\item By using his secret key $sk_b$, decrypt $Y_i$ attribute-wise to get $\gamma_{i,j} = D_{sk_b}(Y_{i,j})$. 
\item Remove randomness from $x_{i,j}$ to get $t'_j = x_{i,j} - \gamma_{i,j}$. Based on the above 
discussions, it is clear that 
$t'$ will be a data record in $D$ that satisfies the input range query $Q$, i.e. $\alpha \le t'_{k} \le \beta$ always 
holds.
\item Finally, Bob adds the data record $t'$ to his output set: $S = S \cup t'$.
\end{itemize}

\subsubsection{Security Analysis}
Informally speaking, during the query request step of Bob, only the additive random 
shares of the boundary values (i.e., $\alpha$ and $\beta$) 
are sent to $C_1$ and $C_2$. That is, $\alpha$ and $\beta$ are never 
revealed to Alice, $C_1$ and $C_2$. 
However, the attribute index $k$ upon 
which he wants to execute the range query is revealed to $C_1$ for efficieny reasons. 
Also, since $C_1$ does not have the decryption key and as all the values 
it receives are in encrypted form, $C_1$ cannot learn 
anything about the original data. In addition, the information $C_2$ has is randomized
by adding randomly chosen numbers. Thus, $C_2$ does not learn anything about the original data either. 
Because each data record is
encrypted attribute-wise, the index $k$, the number of attributes, and the size of the database
do not violate semantic security of the encryption scheme. 
Therefore, the privacy of Bob is always preserved. 

To formally prove the security of PPRQ$_1$ under the semi-honest model, we need to use the Composition Theorem 
given in \cite{goldreich-general}. The theorem says that if a protocol consists of sub-protocols,
the protocol is secure as long as the sub-protocols are secure plus all the intermediate results
are random or pseudo-random. Using the same proof strategies presented in Section \ref{security-scp},
we can easily show that the messages seen by $C_1$ and $C_2$ during 
steps 2, 3, 5 and 6 of Algorithm \ref{alg:pprq1} are pseudo-random values.
In addition, as proved earlier, the \scp~scheme is secure, and the SMP protocol given 
in the Appendix is secure since all the intermediate values are computationally 
indistinguishable from random values. Using the Composition Theorem, we can claim
PPRQ$_1$ is secure under the semi-honest model. In a similar fashion, 
by utilizing the \scp~and SMP protocols that are secure against malicious 
adversaries, we can construct a PPRQ$_1$ protocol that is secure 
under the malicious model.

In the PPRQ$_1$ protocol, the data access 
patterns are protected from both $C_1$ and $C_2$. First, although the outputs 
of \scp~and SMP are revealed to $C_1$, they are in encrypted format.  
Therefore, the data access patterns are protected from $C_1$. In addition, 
even though the vector $Z$ is revealed to $C_2$, it cannot trace 
back to the corresponding data records due to the random permutation of $O$ by $C_1$. Thus, 
the data access patterns are further protected from $C_2$. Also, due to randomization by $C_1$, 
contents of $D$ are never disclosed to $C_2$. However, we emphasize that the value of 
$k$ (part of $Q$) is revealed to $C_1$ for efficiency reasons. Also, $C_2$ will know 
the size of the output set $|S|$, i.e., the number of data 
records satisfying the input range query $Q$. At this point, we believe that 
$|S|$ can be treated as minimal information as it will not be helpful for $C_2$ to 
deduce any information regarding $\alpha, \beta$, and contents of $D$. Hence, 
we claim that the  PPRQ$_1$ protocol preserves the semantic security of the underlying encryption scheme.

\subsubsection{Computation Complexity} In the proposed PPRQ$_1$ protocol, for each record $t_i$, $C_1$ 
and $C_2$ jointly execute \scp~and SMP as sub-routines twice 
and $w+1$ times, respectively. Also, $C_1$ has to randomize the 
attribute values of each record (which requires $w$ encryptions). In addition, he/she has 
to encrypt the corresponding 
random values using Bob's public key. This requires $w$ encryptions per record. Furthermore, $C_2$ has to perform 
$w$ decryptions for each output record. Therefore, for $n$ records, the computation cost of the federated cloud (i.e., 
the combined cost of $C_1$ and $C_2$) is bounded by $O(n)$ instantiations 
of \scp, $O(w*n)$ instantiations of SMP, ~and $O(w*n)$ encryptions (assuming that the encryption and decryption 
times are almost the same under Paillier's scheme).
 
On the other hand, Bob's computation cost mainly depends on the
data decryption step in PPRQ$_1$ in which he has to perform $w$ 
decryptions for each record in $S$. Hence, Bob's 
total computation cost in PPRQ$_1$ is bounded by $O(w*|S|)$ 
encryptions (under the assumption that time for encryption and decryption are the same under Paillier's  scheme). 
Plus, assuming the constant-round \scp~protocol, we claim that PPRQ$_1$ is also bounded 
by a constant number of rounds. For large values of 
$|S|$ (which depends on the query $Q$ and database $D$), Bob's computational 
cost can be high. Therefore, with the goal of improving Bob's efficiency, 
we present an alternate PPRQ protocol in the next sub-section.

\subsection{Protocol 2} 
Similar to PPRQ$_1$, the proposed second protocol (referred to as PPRQ$_2$) consists of 
two cloud providers $C_1$ and $C_2$ where Alice outsources 
her encrypted database to $C_1$. However, unlike PPRQ$_1$, there is no 
need for Bob to generate a public-secret key pair in PPRQ$_2$. Instead, we assume that 
Alice shares her secret key $sk$ between $C_1$ and $C_2$ using threshold-based 
(Paillier) cryptosystem\cite{damgard-2001-thres}. More specifically, let $sk_1$ and $sk_2$ be the shares 
of $sk$ such that Alice sends $sk_1$ and $sk_2$ to $C_1$ and $C_2$, respectively. By doing 
so, PPRQ$_2$ aims at shifting the total expensive operations obliviously between the two clouds; thereby, improving the 
efficiency of Bob in comparison to that of in PPRQ$_1$. That is, the user Bob in PPRQ$_2$ 
can take full advantage of cloud computing at the expense of additional cost on the federated cloud. 
Note that, under the above threshold cryptosystem\cite{damgard-2001-thres}, a decryption operation requires the participation 
of both parties. 
\begin{algorithm}[!htbp]
\begin{algorithmic}[1]
\REQUIRE $sk$ is private to Alice; $sk_1$ and $\pi$ are 
private to $C_1$; $sk_2$ is private to $C_2$; $Q = \{k, \alpha, \beta\}$ is private to Bob\\
\hspace*{-0.4cm}Steps 1 to 4 are the same as in  PPRQ$_1$\\
\setalglineno{5}
\STATE $C_1$:
\begin{enumerate}\itemsep=-1pt    
    \item[(a).] \textbf{for} $1 \leq i \leq n$ and $1 \le j \le w$~\textbf{do:}
              \begin{itemize}\itemsep=-1pt
                \item $U_{i,j} \gets T'_{i,j}\ast E_{pk}(r_{i,j})$, where $r_{i,j} \in_R \mathbb{Z}_N$                 
                \item $H_{i,j} \gets E_{pk}(r_{i,j})$ 
              \end{itemize} 
    \item[(b).] $O'_i \gets D_{sk_1}(O_i)$, for $1 \le i \le n$    
    \item[(c).] Row-wise permutation: $X \gets \pi(U)$ and  $W \gets \pi(H)$
    \item[(d).] $Z \gets \pi(O')$; send $X, W$ and $Z$ to $C_2$
\end{enumerate}
\STATE $C_2$, \textbf{for} $1 \leq i \leq n$ \textbf{do}:
\begin{enumerate}\itemsep=-1pt
    \item[(a).]  \textbf{if} $D_{sk_2}(Z_i)=1$ \textbf{then:}
       \begin{itemize}\itemsep=0pt
          \item \textbf{for} $1 \le j \le w$ \textbf{do}:
         \begin{itemize}\itemsep=0pt
             \item $X'_{i,j} \gets X_{i,j}*E_{pk}(r'_{i,j})$, where $r'_{i,j} \in_R \mathbb{Z}_{N}$ 
             \item $Y'_{i,j} \gets W_{i,j}*E_{pk}(r'_{i,j})$ 
             \item $W'_{i,j} \gets D_{sk_2}(Y'_{i,j})$
          \end{itemize} 
         \item Send $(X'_i,W'_i)$ to $C_1$               
       \end{itemize}  
       \textbf{else}~~  Ignore $(X_i,W_i)$           
\end{enumerate}
\STATE $C_1$, \textbf{foreach} received entry $(X'_i, W'_i)$ from $C_2$ \textbf{do}:
\begin{enumerate}\itemsep=0pt       
    \item[(a).] \textbf{for} $1 \le j \le w$ \textbf{do}:
       \begin{itemize}\itemsep=1pt
                 \item $h_{i,j} \gets D_{sk_1}(W'_{i,j})$
                 \item $H_{i,j} \gets X'_{i,j}*E_{pk}(N - h_{i,j})$   
                 \item $H'_{i,j} \gets H_{i,j} \ast E_{pk}(\hat{r}_{i,j})$, where $\hat{r}_{i,j} \in_R \mathbb{Z}_N$
                 \item $\Phi_{i,j} \gets D_{sk_1}(H'_{i,j})$ ; send $\Phi_{i,j}$ to $C_2$ and $\hat{r}_{i,j}$ to Bob 
       \end{itemize}                
 \end{enumerate}
\STATE $C_2$, \textbf{foreach} received entry $\Phi_i$ from $C_1$ \textbf{do}:
\begin{enumerate}\itemsep=-1pt     
     \item[(a).] \textbf{for} $1 \le j \le w$ \textbf{do}:
       \begin{itemize}                
                \item $ \Gamma_{i,j} \gets D_{sk_2}(\Phi_{i,j})$; send $\Gamma_{i,j}$ to Bob 
       \end{itemize}           
 \end{enumerate}
\STATE Bob:
\begin{enumerate}\itemsep=-1pt
    \item[(a).] $S\gets \emptyset$
    \item[(b).] \textbf{foreach} received entry $(\Gamma_i, \hat{r}_i)$ \textbf{do}:
      \begin{itemize} \itemsep=0pt                
         \item $t'_j \gets \Gamma_{i,j} - \hat{r}_{i,j} \bmod N$, for $1 \le j \le w$           
         \item $S \gets S \cup t'$  
      \end{itemize}
\end{enumerate}
\end{algorithmic}
\caption{$\textrm{PPRQ}_2(T, Q) \rightarrow S$}
\label{alg:pprq2}
\end{algorithm}

We emphasize that the building blocks utilized in this paper, i.e., \scp~and 
SMP, can be easily extended to the threshold-based setting with 
the same security guarantee and outputs. Without loss generality, let \tscp~and TSMP denote the corresponding 
protocols constructed for \scp~and SMP under the threshold-based setting. 

The main steps involved in the proposed PPRQ$_2$ protocol are highlighted in Algorithm \ref{alg:pprq2}.
To start with, upon receiving Bob's query request, $C_1$ and $C_2$ involve 
in the \tscp~and TSMP protocols to compute $O$ and $T'$. This process is similar to steps 1 to 4 of  
PPRQ$_1$. Note that, at the end of this step, only $C_1$ knows $O$ and $T'$. 
After this, $C_1$ randomizes the entries of $T'$ attribute-wise and also encrypts 
the corresponding random factors using the public key $pk$. That is, 
he/she computes $U_{i,j} = T'_{i,j} *E_{pk}(r_{i,j})$ and $H_{i,j}= E_{pk}(r_{i,j})$, 
for $1 \le i \le n$ and $1 \le j \le w$, where $r_{i,j}$ is a random 
number in $\mathbb{Z}_N$. Also, $C_1$ partially decrypts $O$ component-wise using his/her 
secret key share $sk_1$ to get $O'_i = D_{sk_1}(O_i)$, for $\le i \le n$. Then, $C_1$ performs a row-wise permutation 
on $U$ and $H$ to get $X = \pi(U)$ and 
$W = \pi(H)$, respectively. Here $\pi$ is a random permutation function 
known only to $C_1$. In addition, $C_1$ randomly permutes the vector $O'$ to get 
$Z = \pi(O')$. Then, $C_1$ sends $X, W,$ and $Z$ to $C_2$. 

Upon receiving, $C_2$ filters the entries of $(X, W)$ using $Z$ and proceeds 
as follows:
\begin{itemize}
\item Decrypt each entry in $Z$ using his/her secret key share $sk_2$ and check whether it is 
0 or 1. Similar to PPRQ$_1$, if $D_{sk_2}(Z_i) = 1$, then we observe that the corresponding data record  
$X_i$ satisfies the range query condition. Under this case, $C_2$ randomizes both $X_i$ and $W_i$ 
attribute-wise using Alice's public key $pk$. More specifically, 
he/she computes $X'_{i,j} = X_{i,j}*E_{pk}(r'_{i,j})$ and 
$Y'_{i,j} = W_{i,j}*E_{pk}(r'_{i,j})$, where $r'_{i,j}$ is a random number 
in $\mathbb{Z}_N$ known only to $C_2$.  Then, $C_2$ partially decrypts $Y'_{i,j}$ to get 
$W'_{i,j} \gets D_{sk_2}(Y'_{i,j})$ and sends $(X'_i, W'_i)$ to $C_1$. 
\item On the other hand, if $D_{sk_2}(Z_i) = 0$, then the corresponding data record $X_i$ do not 
satisfy the query condition. Therefore, $C_2$ simply ignores $(X_i, W_i)$. 
\end{itemize}
Now, for each received entry $(X'_i, W'_i)$, $C_1$ performs the following 
set of operations to compute the encrypted versions of data records that 
satisfy the query condition locally:
\begin{itemize}
\item Decrypt $W'_i$ attribute-wise using his/her secret key share $sk_1$. That is, compute 
$h_{i,j} =  D_{sk_1}(W'_{i,j})$. Observe that $h_{i,j} = r_{i,j} + r'_{i,j} \bmod N$, where 
$1 \le j \le w$ and $r'_{i,j}$ is known only to $C_2$. 
\item Remove the random factors (within the encryption) from $X'_i$ attribute-wise by computing 
$H_{i,j} = X'_{i,j}*E_{pk}(N - h_{i,j})$. Note that $N-h_{i,j}$ is equivalent to $-h_{i,j}$ under 
$\mathbb{Z}_N$. By the end of this step, $C_1$ has encrypted data records 
$H_{i}$ that satisfy Bob's range query.
\end{itemize}
Now, $C_1$ randomizes $H_i$ attribute-wise to get $H'_{i,j} = H_{i,j}*E_{pk}(\hat{r}_{i,j})$, for $1 \le j \le w$.  
Here $\hat{r}_{i,j}$ is a random number in $\mathbb{Z}_N$ known only to $C_1$. 
Also, $C_1$ partially decrypts $H'_i$ attribute-wise to get $\Phi_{i,j} = D_{sk_1}(H'_{i,j})$, 
sends $\Phi_{i,j}$ to $C_2$ and $\hat{r}_{i,j}$ to Bob, for $1 \le j \le w$. 

In addition, for each received entry $\Phi_{i}$, $C_2$ decrypts it attribute-wise 
to get $\Gamma_{i,j} = D_{sk_2}(\Phi_{i,j})$ and sends the results to Bob. Note that, due to randomization 
by $C_1$, $\Gamma_{i,j}$ is always a random number in $\mathbb{Z}_N$.

Finally, for each 
received entry pair $(\Gamma_i, \hat{r}_i)$, Bob retrieves 
the corresponding output record and proceeds as below:
\begin{itemize}\itemsep=0pt
\item Remove randomness from $\Gamma_i$ attribute-wise to 
get $t'_i = \Gamma_{i,j} - \hat{r}_{i,j} \bmod N$, for $1 \le j \le w$. We observe that $t'\in D$ and the 
property $\alpha \le t'_k\le  \beta$ always holds.
\item Include data record $t'$ to the output set: $S = S\cup t'$.
\end{itemize}
\subsubsection{Security Analysis}
The security proof of  PPRQ$_2$ is similar to that of PPRQ$_1$. 
Briefly, due to random permutation of $O'$ by $C_1$; 
$C_2$ cannot trace back to the data records satisfying 
the query condition. In addition, as the comparison 
results (in encrypted form) are known only 
to $C_1$ who does not have access to the secret key $sk$, the data access 
patterns are protected from $C_1$. Therefore, we claim that the 
data access patterns are protected from both $C_1$ and $C_2$. Furthermore, no other 
information regarding the contents of $D$ is revealed to the cloud service providers since 
the intermediate decrypted values are random in $\mathbb{Z}_N$.  
However, in PPRQ$_2$, $k$ (part of $Q$) is revealed to $C_1$ whereas 
$|S|$ is revealed to $C_1$ and $C_2$. As mentioned earlier in the security analysis 
of PPRQ$_1$, this is treated 
as a minimal information leakage since it cannot be used to break
the semantic security of the encryption scheme. 

\subsubsection{Computation Complexity} 
The computation cost of the federated cloud 
(i.e., the combined cost of $C_1$ and $C_2$) in PPRQ$_2$ is 
bounded by $O(n)$ instantiations of \tscp, $O(w*n)$ instantiations of TSMP  
and $O(w*(n+ |S|))$ encryptions and decryptions. In general, assuming the decryption time 
under threshold cryptosystem is (at most) two times more than an encryption operation, 
the computation cost of the federated cloud in PPRQ$_2$ is (at most) twice to that of PPRQ$_1$. 
However, unlike PPRQ$_1$, during the data retrieval step of 
PPRQ$_2$, Bob does not perform any decryption operations. Thus, the computation cost of 
Bob in PPRQ$_2$ is negligible compared that of in PPRQ$_1$. 
Remember that Bob's computation cost in the PPRQ$_1$ protocol 
is bounded by $O(w*|S|)$ decryptions. 

At first, it seems that the proposed PPRQ protocols are costly and may not scale 
well for large databases. However, we stress 
that the computations involved on each data record are fully independent of others. 
In particular, the execution of sub-routines \scp~and SMP (similarly, \tscp~and TSMP) on 
a data record does not depend on the operations of other data records. Therefore, in the 
cloud computing environment where high performance parallel processing 
can be easily achieved using multiple cores, we believe that 
the scalability issue in the proposed PPRQ protocols can be eliminated or mitigated. Furthermore, by 
using the existing MapReduce techniques (such as Hadoop\cite{hadoop-2009}) in the cloud, the performance 
of the proposed PPRQ protocols can be improved drastically. We leave 
the above low-level implementation details for future work. Nevertheless, the main advantages 
of the proposed PPRQ protocols are that they protect data confidentiality and privacy 
of user's input query. In addition, they protect data access patterns and in particular 
PPRQ$_2$ incurs 
negligible computation cost on the end-user.


\section{Conclusions}\label{sec:concl}
Query processing in distributed databases has been well-studied in the literature.   
In this paper, 
we focus on the privacy-preserving range query (PPRQ) problem over encrypted data 
in the cloud. We observed that most of the existing PPRQ methods reveal valuable information, such as 
data access 
patterns, to the cloud provider; thus, they are not secure from both
data owner and query issuer's perspective. 

In general, the basic security primitive that is required to solve 
the PPRQ problem is the secure comparison (SC) of encrypted integers. Since the 
existing SC methods (both custom-designed and garbled-circuit approaches) 
are not efficient, we first proposed a new probabilistic SC scheme 
that is more efficient than the current state-of-the-art SC protocols. Then, we proposed two novel 
PPRQ protocols by using our SC scheme as the building block under the cloud computing environment. Besides 
ensuring data confidentiality, the proposed PPRQ protocols protect 
query privacy and data access patterns from the cloud service providers. 
In addition, from end-user's perspective, our second protocol is significantly more efficient than our first protocol. 

In this work, we proposed the SC scheme whose round 
complexity is bounded by $O(m)$. Therefore, developing a constant round 
SC protocol using Carry-Lookahead Adders will be the primary focus of our future work. 
Another interesting direction is to extend our PPRQ protocols to multi-dimensional range queries and analyze 
their trade-offs between security and efficiency. We will also 
investigate alternative methods and extend our work to other complex conjunctive 
queries.

\bibliographystyle{abbrv}
\bibliography{ref}

\begin{center} \Large{\textbf{Appendix}}\end{center}
\textbf{Possible Implementation of SMP. } 
Consider a party $P_1$ with private input $(E_{pk}(a), E_{pk}(b))$ and a 
party $P_2$ with the secret key $sk$. The goal of the secure multiplication (SMP) 
protocol is to return the encryption 
of $a \ast b$, i.e., $E_{pk}(a*b)$ as the output to $P_1$. During this protocol, no information regarding 
$a$ and $b$ should be revealed to $P_1$ and $P_2$. First, we emphasize that one can 
construct a SMP protocol by using the garbled-circuit technique. However, we 
observe that our custom-designed SMP 
protocol (as explained below) is more efficient than the circuit-based method. The basic idea of our SMP protocol 
is based on the following property which holds 
for any given $a,b \in \mathbb{Z}_N$: 
\begin{equation}\label{eq:mult}
 a\ast b = (a+r_a)\ast (b+r_b) - a\ast r_b - b\ast r_a - r_a\ast r_b
\end{equation}
\noindent where all the arithmetic operations are performed under $\mathbb{Z}_N$. The overall 
steps involved in the proposed SMP protocol are shown in Algorithm \ref{alg:sm}. 
Briefly, $P_1$ initially randomizes $a$ and $b$ by computing $a' = E_{pk}(a)*E_{pk}(r_a)$ and $b' = E_{pk}(b)*E_{pk}(r_b)$, and 
sends them to $P_2$. Here $r_a$ and $r_b$ are random numbers in $\mathbb{Z}_N$ known only to $P_1$. 
Upon receiving, $P_2$ decrypts and multiplies them to get $h = (a+r_a)\ast(b+r_b) \bmod N$. 
Then, $P_2$ encrypts $h$ and sends it to $P_1$. After this, $P_1$ removes extra random factors 
from $h' = E_{pk}((a+r_a)*(b+r_b))$ based on Equation \ref{eq:mult} to get $E_{pk}(a*b)$. 
Note that, under Paillier cryptosystem,  ``$N-x$'' is equivalent to ``$-x$'' in $\mathbb{Z}_N$.

\begin{algorithm}[h]
\begin{algorithmic}[1]
\REQUIRE $P_1$ has $E_{pk}(a)$ and $E_{pk}(b)$; $P_2$ has $sk$
\STATE $P_1$:
\begin{enumerate}\itemsep=0pt
    \item[(a).]  Pick two random numbers $r_a, r_b \in \mathbb{Z}_N$
    \item[(b).]  $a' \gets E_{pk}(a)\ast E_{pk}(r_a)$
    \item[(c).]  $b' \gets E_{pk}(b)\ast E_{pk}(r_b)$; send $a', b'$ to $P_2$    
\end{enumerate}
\STATE $P_2$:
\begin{enumerate}\itemsep=0pt
    \item[(a).]  $h_a \gets D_{sk}(a')$;~ $h_b \gets D_{sk}(b')$
    \item[(b).] $h \gets h_a \ast h_b \bmod N$
    \item[(c).] $h' \gets E_{pk}(h)$; send $h'$ to $P_1$
\end{enumerate}
\STATE $P_1$:
\begin{enumerate}\itemsep=0pt
    \item[(a).]  $s \gets h' \ast E_{pk}(a)^{N- r_b}$
    \item[(b).]  $s' \gets s \ast E_{pk}(b)^{N- r_a}$
    \item[(c).]  $E_{pk}(a\ast b) \gets s'\ast E_{pk}(r_a\ast r_b)^{N-1}$
\end{enumerate}
\end{algorithmic}
\caption{SMP$(E_{pk}(a), E_{pk}(b)) \rightarrow E_{pk}(a\ast b)$}
\label{alg:sm}
\end{algorithm}

\end{document}